\documentclass[twocolumn]{aastex631} 
\usepackage{amsmath, amsthm}
\allowdisplaybreaks[1]

\submitjournal{ApJL}

\shorttitle{Tidal Disruption Events from Stripped Stars}
\shortauthors{Mockler et al.}
\graphicspath{{./}{figures/}}

\begin{document}

\title{Tidal Disruption Events from Stripped Stars}


\correspondingauthor{Brenna Mockler}
\email{bmockler@carnegiescience.edu}

\author[0000-0001-6350-8168]{Brenna Mockler}
\affiliation{The Observatories of the Carnegie Institution for Science, Pasadena, CA 91101, USA}

\author[0000-0003-0648-2402]{Monica Gallegos-Garcia}

\affiliation{Department of Physics \& Astronomy, Northwestern University, Evanston, IL 60208, USA}
\affiliation{Center for Interdisciplinary Exploration \& Research in Astrophysics (CIERA), Northwestern University, Evanston, IL 60208, USA}
\author[0000-0002-6960-6911]{Ylva G\"{o}tberg}
\affiliation{Institute of Science and Technology Austria, 3400 Klosterneuburg, Austria}
\affiliation{The Observatories of the Carnegie Institution for Science, Pasadena, CA 91101, USA}

\author[0000-0003-2869-7682]{Jon M. Miller}
\affiliation{Department of Astronomy, University of Michigan, 1085 South University Avenue, Ann Arbor, MI 48109, USA}

\author[0000-0003-2558-3102]{Enrico Ramirez-Ruiz}
\affiliation{Department of Astronomy and Astrophysics, University of California, Santa Cruz, CA 95064, USA}

\begin{abstract}
Observations of tidal disruption events (TDEs) show signs of Nitrogen enrichment reminiscent of other astrophysical sources such as active galactic nuclei (AGN) and star-forming galaxies. Given that TDEs probe the gas from a single star, it is possible to test if the observed enrichment is consistent with expectations from the CNO cycle by looking at the observed Nitrogen/Carbon (N/C) abundance ratios. Given that $\approx 20\%$ of solar mass stars (and an even larger fraction of more massive stars) live in close binaries, it is worthwhile to also consider what TDEs from stars influenced by binary evolution would look like. We show here that TDEs from stars stripped of their Hydrogen-rich (and Nitrogen-poor) envelopes through previous binary-induced mass loss can produce much higher observable N/C enhancements than even TDEs from massive stars. Additionally, we predict that the time-dependence of the N/C abundance ratio in the mass fallback rate of stripped stars will follow the inverse behavior of main-sequence stars, enabling a more accurate  characterization of the disrupted star.
\end{abstract}


\section{Introduction} \label{sec:intro}
With recent observations and new, more sensitive instruments, Nitrogen enrichment has been appearing in a wide variety of astrophysical research: from quasars and active galactic nuclei (AGN) disks \citep[e.g.][]{bentz_nitrogen-enriched_2004} to distant star-forming galaxies \citep[e.g.][]{bunker_jades_2023}. Nitrogen enrichment is typically closely linked to the CNO cycle -- the nuclear reaction that fuses Hydrogen to Helium in the deep interiors of massive stars. How Nitrogen enriched material is then exposed from its origin in the interior of massive stars remains a largely unsolved problem. This emphasizes the importance of understanding links between stellar evolution and a range of different astrophysical research fields and in a variety of different environments.

Recent observations of tidal disruption events (TDEs) show surprisingly high N/C abundance ratios (e.g. abundance ratios relative to solar of N/C $>10$ in ASASSN-14li, iPTF15af, iPTF16fnl, see \citealt{cenko_ultraviolet_2016, blagorodnova_iptf16fnl_2017, blagorodnova_broad_2019, Kochanek:2016a, Yang:2017})\footnote{In addition to these three events, there are also a few newer TDEs with potential detections of broad UV Nitrogen and/or Carbon lines that might be used to constrain N/C ratios with further analysis. These include ASASSN-14ko \citep{payne_chandra_2023}, AT2018zr \citep{hung_discovery_2019}, AT2019qiz \citep{hung_discovery_2021}, and AT2020vdq \citep{somalwar_first_2023}.}. These TDEs occur when stars are ripped apart by the supermassive black holes (SMBHs) in the centers of galaxies, after which the gas from the star is eventually accreted by the SMBH, producing bright transient flares in the process \citep[e.g.][]{Rees:1988a, Evans:1989a,Guillochon:2013a}. High N/C abundance ratios suggest the disrupted star was subject to the CNO cycle and therefore relatively massive, thus constituting a promising channel for enriching gas surrounding the SMBH \citep{mockler_evidence_2022}. With the increasing sample of TDEs now reaching the $\gtrsim 100$ \citep[e.g.][]{hammerstein_final_2023}, and with upcoming data from the Vera Rubin Observatory predicted to increase this number to $\gtrsim 1000$ \citep[e.g.][]{Bricman:2020}, we can expect a wealth of TDE data soon to help improve our understanding not only of SMBHs, but also stellar populations 
in the centers of galaxies.

Tidal disruptions of stars around supermassive black holes (SMBHs) probe stellar populations on size scales that cannot be observed directly outside our own galactic neighborhood. The stars that are disrupted originate from near the sphere of influence of the black hole, corresponding to size scales of order $\approx 0.5 - 10$pc for SMBHs with masses between $\approx 10^5 - 10^8 M_\odot$.   

The increasing number of TDE observations has motivated the development of numerous light curve and spectral models to attempt to explain the unique features of these events and also to use observations to constrain properties of the black hole and the disrupted star \citep{Roth:2018, mockler_weighing_2019, ryu_measuring_2020, wen_continuum-fitting_2020, metzger_cooling_2022} and better understand black hole accretion \citep{Ramirez-Ruiz:2009a, Dai:2018, andalman_tidal_2022}. 

Much like in observations of AGN \citep[e.g.][]{bentz_nitrogen-enriched_2004,jiang_sample_2008}, measurements of spectral lines in TDEs hold the potential to constrain the composition of the irradiated gas in galactic nuclei. Unlike in AGN, in TDEs the gas producing the lines is predominantly from a single star, and so composition constraints in TDE spectra tell us about the properties of individual stars in the spheres of influence of the host galaxies.

The observations of TDEs with unexpectedly high Nitrogen abundances have complicated the previous assumption that most TDEs originated from low-mass main-sequence stars (which are generically Nitrogen-poor), as is predicted if disrupted stars are drawn randomly from the stellar population given standard stellar initial mass functions. \citet{mockler_evidence_2022} showed that massive stars ($M>2 \, M_{\odot}$) that fuse Hydrogen through the CNO cycle are somewhat more consistent with these observations, since the abundance of Nitrogen rises at expense of Carbon and Oxygen in regions that experience the CNO cycle. However, recent re-analysis of X-ray data from the TDE ASASSN-14li by \citet{miller_evidence_2023} find Nitrogen/Carbon abundances that push the limits of what is expected even from the disruption of massive stars. Main-sequence stars have Nitrogen-rich cores, but their envelopes are Nitrogen-poor. During a TDE, dilution by in-mixing of the Nitrogen-poor envelope should make the abundance ratio less pronounced. 

Motivated by these results, in this work we explore what the disruption of a star stripped of its Hydrogen-rich (and therefore Nitrogen-poor) envelope would look like and how it would compare to the disruption of stars at various stages of stellar evolution. Such ``stripped stars'' are expected to be created through binary interaction over the full stellar mass range, but we explore stripped stars originating from massive stellar evolution as they will result in higher N/C abundance ratios. Throughout this work, we will use N/C to mean the abundance ratio of Nitrogen to Carbon relative to the solar value\footnote{Note that the mass abundance ratio relative to solar is equivalent to the number abundance ratio relative to solar as the ratio of the mass of the elements cancels out.}. 

\section{The compositional structure of stripped stars}\label{sec:comp}

\begin{figure}[ht!]
\begin{center}
\includegraphics[width = \columnwidth]{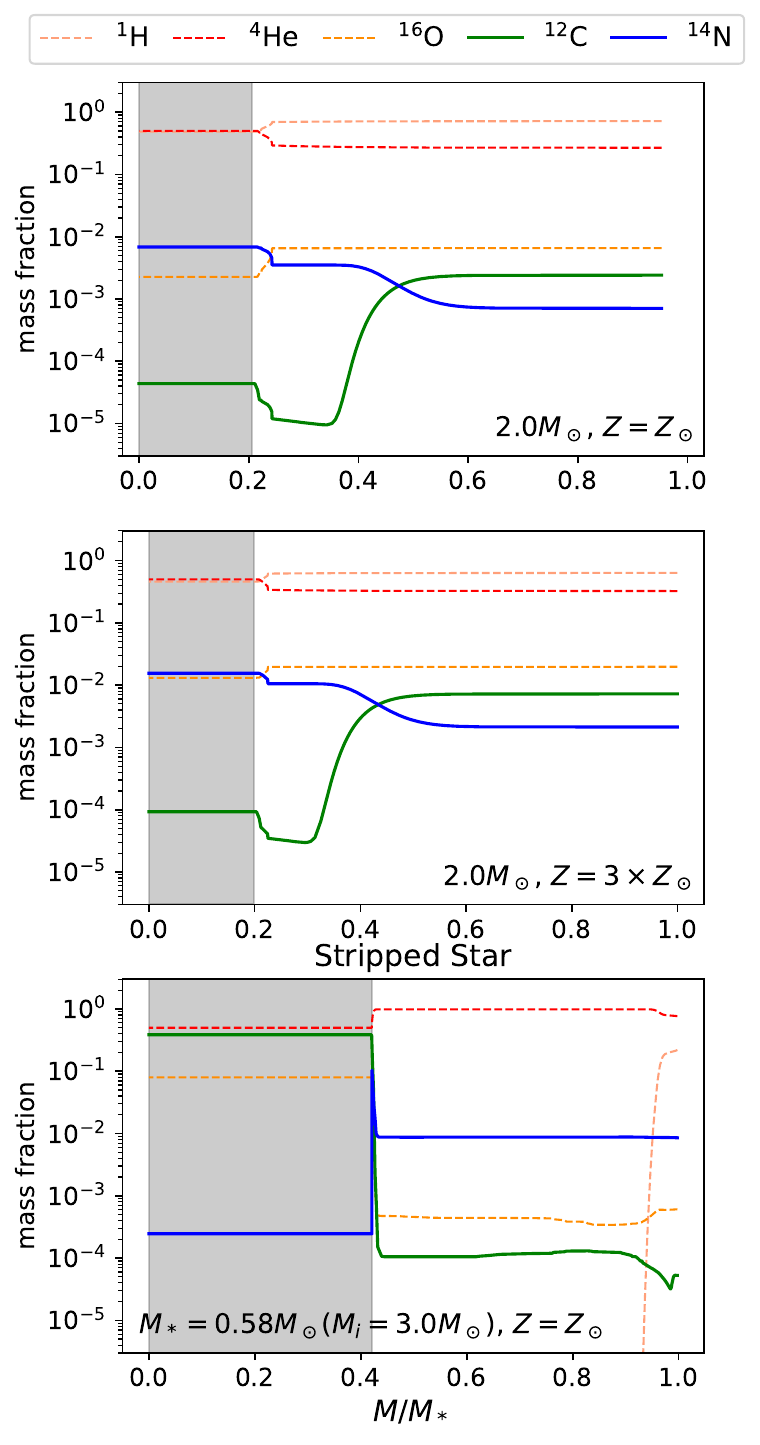} 
\end{center}
\vspace{-0.5cm}
\caption{Composition as a function of mass fraction for three representative stars. The top plot is of a $2 M_\odot$ MS star born at $Z_\odot$, the middle plot is of a $2 M_\odot$ MS star born at $3 Z_\odot$, and the bottom plot is of a stripped star born at $Z_\odot$ with an initial mass of $3 M_\odot$ and a current mass of $0.58 M_\odot$. All three stars have $Y_c = 0.5$. The two MS stars are halfway through Hydrogen burning, and the stripped star is halfway through Helium burning. The gray shaded regions denote the convectively burning core (including convection and convective overshoot).
\label{fig:composition_mass_fraction}
}
\end{figure}

Stripped stars have naturally Nitrogen-enriched and Carbon-depleted outer layers \citep{gotberg_spectral_2018}. This originates from the previous main-sequence evolution when Hydrogen fused to Helium through the CNO cycle. The slowest reaction in the CNO cycle is proton capture onto Nitrogen, which causes a build-up of Nitrogen at the expense of Carbon and Oxygen in regions that have experienced or are experiencing the CNO cycle \citep[e.g.][]{burbidge_synthesis_1957}. As the convective main-sequence {\it core} recedes in mass coordinate, seen from outside and inwards, it leaves a layer that is gradually enriched in Helium, but that also has Nitrogen-enrichment from the CNO cycle throughout it. During envelope-stripping in binaries, the thick Hydrogen-rich (and Nitrogen-poor) {\it envelope} is removed, revealing the chemically enriched layer. Because of this, the surfaces of stripped stars contain a small amount of Hydrogen \citep[notably still sufficient to produce strong Hydrogen lines in supernova,][]{dessart_core-collapse_2011, dessart_nature_2012}, a large amount of Helium, and are enriched in Nitrogen \citep[][see also \citealt{drilling_mk-like_2013}]{hirsch_hot_2009, heber_hot_2016, gotberg_stellar_2023}.  

After envelope-stripping is complete, the stripped star starts fusing Helium to Carbon and Oxygen in the center \citep{laplace_different_2021}. The central region is convective, but it is relatively small and there is also a substantial radiative layer that previously was the main-sequence core and now extends to the stellar surface. The described composition structure is depicted in Figure~\ref{fig:composition_mass_fraction}, where we show a model for a $0.58 M_\odot$ stripped star (with initial mass $M_i = 3M_\odot$, bottom panel) compared with models for $2M_\odot$ main-sequence stars born at two different metallicities that are undergoing CNO cycle Hydrogen burning (top, middle). In this study, we consider the gray shaded (convective) regions the core, and the white (radiative) regions the envelope.  The main-sequence stars are born at solar metallicity ($Z_\odot$, top panel) and $3\times$ solar metallicity ($3 Z_\odot$, middle panel). In Figure~\ref{fig:composition_mass_fraction}, all three models are plotted when the mass fraction of Helium at the center of the star ($Y_c$)$= 0.5$. The main sequence (MS) models are $\sim$halfway though Hydrogen burning, and the stripped star is $\sim$halfway through Helium burning\footnote{This is equivalent to $0.47\times$ and $0.4 \times$ the terminal age main sequence (TAMS) lifetimes respectively for the $Z_\odot$ and $3 Z_\odot$ MS models, and to $0.5\times$ the stripped star's Helium burning lifetime for that model.}. All models were computed with the MESA stellar evolution code \citep{paxton_modules_2011, paxton_modules_2013, paxton_modules_2015, paxton_modules_2018, paxton_modules_2019, jermyn_modules_2023} and the stripped star model was published in \citep[][see their Table 1]{gotberg_spectral_2018}.
We have chosen these masses for analysis because they are on the lower end of the mass range that experience CNO burning, and are therefore representative of the most common stars with Nitrogen enhancements from CNO burning\footnote{We choose this $3 M_\odot$ model because the structure was numerically robust at smaller scales. The qualitative features are the same for a stripped star with a $2M_\odot$ progenitor.} We will continue to use these same three stellar models throughout the paper (at various stages of their evolution). 

To summarize, the entirety of stripped stars are made up of material that has previously experienced CNO burning, however their convective cores are currently undergoing Helium burning. Because of this, their N/C abundance ratios are high in their radiative outer regions (leftover from their MS core CNO burning) but low in their convective cores where Helium burning is ongoing. This is in contrast to MS stars sufficiently massive to experience CNO burning, whose Nitrogen-rich cores are embedded within Nitrogen-poor outer layers. Therefore, to first order, the abundances in the center versus the outer layers of stripped stars exhibit the opposite trends compared to solar type main-sequence stars (see Figure~\ref{fig:composition_mass_fraction}). As we will show, this means that the time-dependent composition evolution in the material returning to a SMBH after the tidal disruption of stripped stars should follow the inverse behavior of CNO burning main-sequence stars.

\section{Connections to observations}\label{sec:connections}

We begin with a very brief overview of the tidal disruption process and go on to describe how the disruption of a stripped star differs from (and is similar to) that of a main-sequence star. 

Tidal disruption occurs when the binding energy of a star is overwhelmed by the tidal gravity of a black hole. The radius of disruption for a star of mass $M_*$ and radius $R_*$ by a black hole of mass $M_h$ can be approximated by setting the self-gravity of the star equal to the tidal force of the black hole, which gives the canonical tidal radius $R_t = R_* \big(\frac{M_*}{M_h}\big)^{1/3}$. The `depth' of disruption relative to $R_t$ can be described using the impact parameter $\beta  = R_t/R_p$, where deeper (smaller pericenter radius, $R_p$) disruptions have larger values of $\beta$. The rate of return of gas to a SMBH after a star is tidally disrupted depends on the binding energy distribution in the gas \citep[e.g.][]{lodato_stellar_2009, guillochon_hydrodynamical_2013, law-smith_stellar_2020, ryu_tidal_2020}. In a full disruption, where the core is destroyed, material in the outer layers of the star (which is more bound to the black hole), will return first, and material from the core of the star (which is less bound to the black hole), will return later. Once the slope of the binding energy distribution as a function of stellar mass becomes relatively constant (as we approach the core of the star), the mass fallback rate approaches the canonical value of $t^{-5/3}$:


\begin{equation}
    \frac{dm}{dt} = \frac{dm}{de} \times \frac{de}{dt}
\end{equation}
\begin{equation}
       \frac{dm}{dt} = \frac{dm}{de} \times \frac{(2 \pi GM_h)^{2/3}}{3} t^{-5/3}
\end{equation}

\noindent Here, $\frac{dm}{dt}$ is the mass return rate to the black hole, $\frac{dm}{de}$ is the mass-energy distribution in the debris, and $\frac{de}{dt}$ is the derivative of the orbital energy with respect to orbital timescale and orbital energy, which can be calculated analytically from Kepler's laws.

\subsection{Composition evolution}

\begin{figure}[ht!]
\begin{center}
\includegraphics[width = \columnwidth]{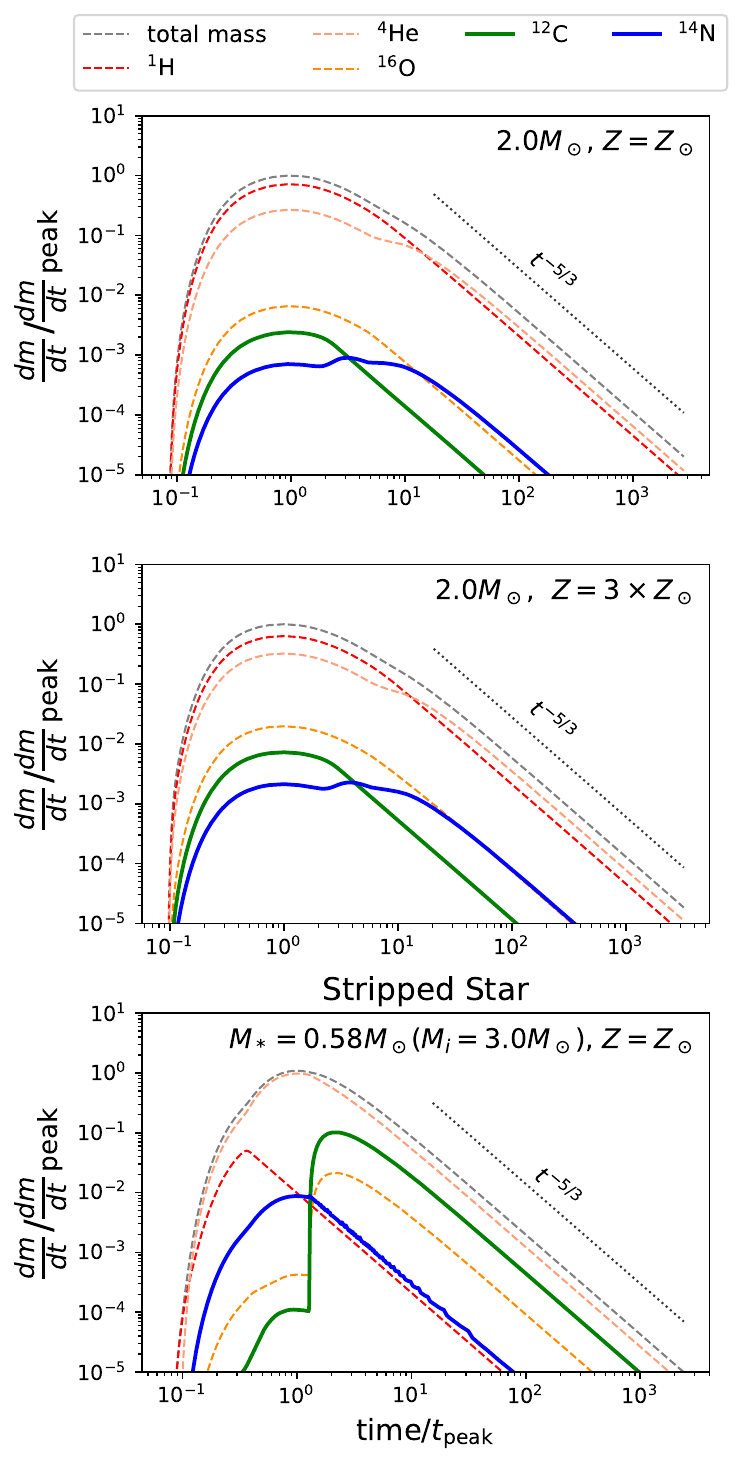} 
\end{center}
\vspace{-0.8cm}
\caption{
We plot analytical fallback rates calculated for various MESA stellar models. Fallback rates for different elements are plotted separately in addition to the overall fallback rate. All fallback rates are scaled by the peak fallback rate and peak timescale for all material (see Figure~\ref{fig:rho_profile} for a comparison with hydro simulations in physical units). The top two panels feature our fiducial $2 M_\odot$ single stars near the end of their MS lifetime at $Y_c = 0.9$. The bottom panel features our $0.58 M_\odot$ stripped star halfway through He-core burning at $Y_c = 0.5$. 
\label{fig:dmdt_composition}
}
\end{figure}

If the star's composition varies with radius (and therefore also with binding energy), this radial stratification transforms into a time-dependence, where the composition of the gas returning to the black hole varies with time. This has been explored for main sequence stars analytically in \citet{Gallegos-Garcia:2018} and with hydrodynamical simulations in \citet{law-smith_tidal_2019}. 

A key result in both is that the ratio of N/C in the fallback material varies dramatically with time and is dependent on the mass and age of the star. As discussed in Section~\ref{sec:comp}, the composition profile of a stripped star is, to first order, the inverse of the profile of a main sequence star (see also Figure~\ref{fig:composition_mass_fraction}). In Figure~\ref{fig:dmdt_composition}, we plot the normalized, time-dependent composition of the fallback rate for our two main sequence stars and compare it to the composition of the fallback rate of our stripped star (of similar progenitor mass). We calculate the fallback rates analytically using the frozen-in approximation described in \citet{Lodato:2009a} and \citet{Kesden:2012b} and apply this framework to MESA stellar models as outlined  in \citet{Gallegos-Garcia:2018}. 

While we know the analytical fallback rates do not exactly match the peak magnitude and timescales of fallback rates from hydrodynamic simulations (and for this reason we scale out these quantities from our composition plots), they should capture overall trends in the composition as a function of time\footnote{see Section~\ref{sec:fallback} and Figure~\ref{fig:rho_profile} for a comparison with fallback curves from simulations}. Hydrodynamic simulations of solar mass main sequence stars have shown a similar transition in the abundances of Nitrogen, Carbon and Oxygen (and the abundance ratio of N/C), in the fallback material of a TDE compared to analytical calculations. This transition corresponds to when material from the core begins to return to the SMBH. Simulations do find that this transition occurs more gradually than analytical treatments predict, and that it begins slightly before peak instead of slightly after peak \citep[]{law-smith_tidal_2019}. Because of this, we expect mixing in the hydrodynamics of disruption to also smear out the sharp transition we see in the relative abundance of these elements in the stripped star (these complications are discussed in more detail in Section~\ref{sec:outflows}).

\begin{figure}[ht!]
\begin{center}
\includegraphics[width = \columnwidth]{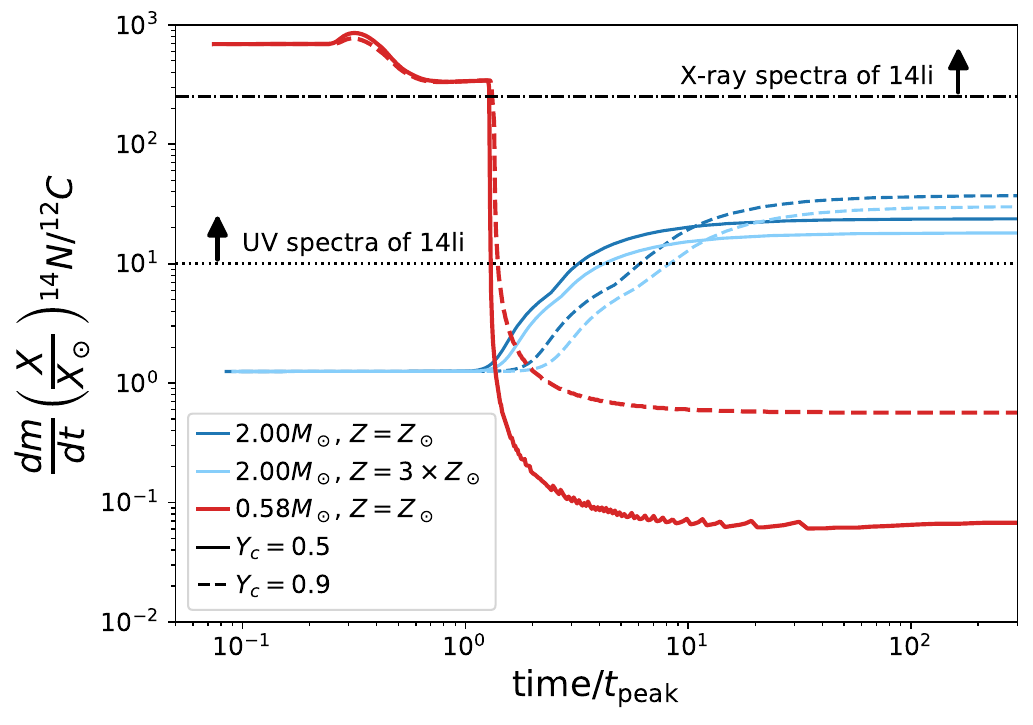}
\end{center}
\vspace{-0.5cm}
\caption{The analytical N/C abundance ratio (relative to solar) in the mass fallback rate as a function of time since disruption (scaled by $t_{\rm peak}$, which we expect to be of order $\sim$ week timescales, see Figure~\ref{fig:rho_profile}). We plot our $0.58 M_\odot$ stripped star ($M_i = 3 M_\odot$) at the beginning of Helium core burning ($Y_c = 0.9$, red solid line) and halfway through Helium core burning ($Y_c = 0.5$, red dashed line). We plot our $2 M_\odot$ MS stars partway through the main sequence and nearing the end of the main sequence. The $Z_\odot$ MS model is plotted in dark blue, the $3\times Z_\odot$ MS model is plotted in light blue ($Y_c = 0.5$ plotted in solid lines, $Y_c = 0.9$ plotted in dashed lines). 
Constraints from UV and X-ray spectra from the transient ASASSN-14li plotted as dotted and dash-dotted lines respectively.}
\label{fig:N14_C12}
\end{figure}

We find that the mass fractions of Nitrogen and Carbon vary much more dramatically in the disruption from our stripped star compared to the main sequence stars throughout the fallback evolution (see Figure~\ref{fig:dmdt_composition}). This is despite comparing to main sequence stars near the end of their MS lifetimes with $Y_c = 0.9$\footnote{This is equivalent to $0.94$ and $0.96 \times \rm TAMS$ for the $Z_\odot$ and $3 \times Z_\odot$ stars respectively.}. Naively, we might expect the material returning at late times in a main sequence star disruption (and therefore including material from the core) to be similar to the material that returns at early times for the stripped star (given that the envelope of the stripped star is made up of material that was near the core when it was on the main sequence). However, because tidal disruption is an inherently aspherical process, we are not probing purely radial slices at each point in time. Instead we are probing vertical slices (compared to the orbital plane of the black hole and star), and so, while material that returns at later times does have more material from the core, it is mixed with material from the outer layers as well (see Figure 1 in \citealt{Gallegos-Garcia:2018} for reference). On the other hand, material that returns at early times originates almost completely from the envelope of the star. This means that for a main sequence star, material from the core that returns at late times and is enhanced in N/C will be diluted by material from the outer layers, whereas for a stripped star, material that returns at early times from the envelope and is also enhanced in N/C will not be as diluted and will retain a higher N/C abundance.

We also find that there is comparatively little difference between the relative mass fractions and N/C abundance ratios in the fallback rates of a main sequence star born at solar metallicity and one born at $3\times$ solar metallicity (top two panels of Figure~\ref{fig:dmdt_composition}, light and dark blue lines in Figure~\ref{fig:N14_C12}). The curves for elements other than Hydrogen are shifted up for the higher metallicity star compared to the solar metallicity star (as expected), but the relative abundances are very similar. The N/C abundance ratio is actually slightly higher at late times in the solar metallicity model than the higher metallicity model. This is because the core of the solar metallicity model makes up a slightly larger fraction of the total mass (the N/C ratio in the core is not significantly larger, it is simply slightly less diluted by the outer layers of the star).

In Figure~\ref{fig:N14_C12}, we focus on just the N/C abundance in the fallback material, and compare this time-dependent composition ratio for stellar models at different stages of evolution. We plot the stripped star and the main sequence stars at $Y_c = 0.5$ and $Y_c = 0.9$. For the main sequence stars, this is partway through the main sequence ($Y_c = 0.5$) and near the end of the main sequence ($Y_c = 0.9$). For the stripped star, this is early-on in Helium burning ($Y_c = 0.9$) and halfway through Helium burning ($Y_c = 0.5$)\footnote{This is equivalent to $0.5$ and $0.12$ of the way through the stripped star's Helium burning lifetime.}.  

Notably, we find that the composition of the fallback rate from the stripped star TDE can naturally explain the very high N/C ratios observed in the X-ray spectra of ASASSN-14li (dash-dotted lines in Figure~\ref{fig:N14_C12}). While moderately massive ($M_* \gtrsim 1-2 M_\odot$) main-sequence stars provide a good explanation for the high N/C constraints measured in the UV \citep[]{yang_carbon_2017, mockler_evidence_2022}, the much higher constraints from X-ray measurements are better matched by the stripped star's fallback rate \citep[]{miller_evidence_2023}. We discuss ASASSN-14li as a potential stripped star TDE candidate further in Section~\ref{sec:14li}.

\subsection{Fallback rate}\label{sec:fallback}

\begin{figure}[ht!]
\begin{center}
\includegraphics[scale = 0.55]{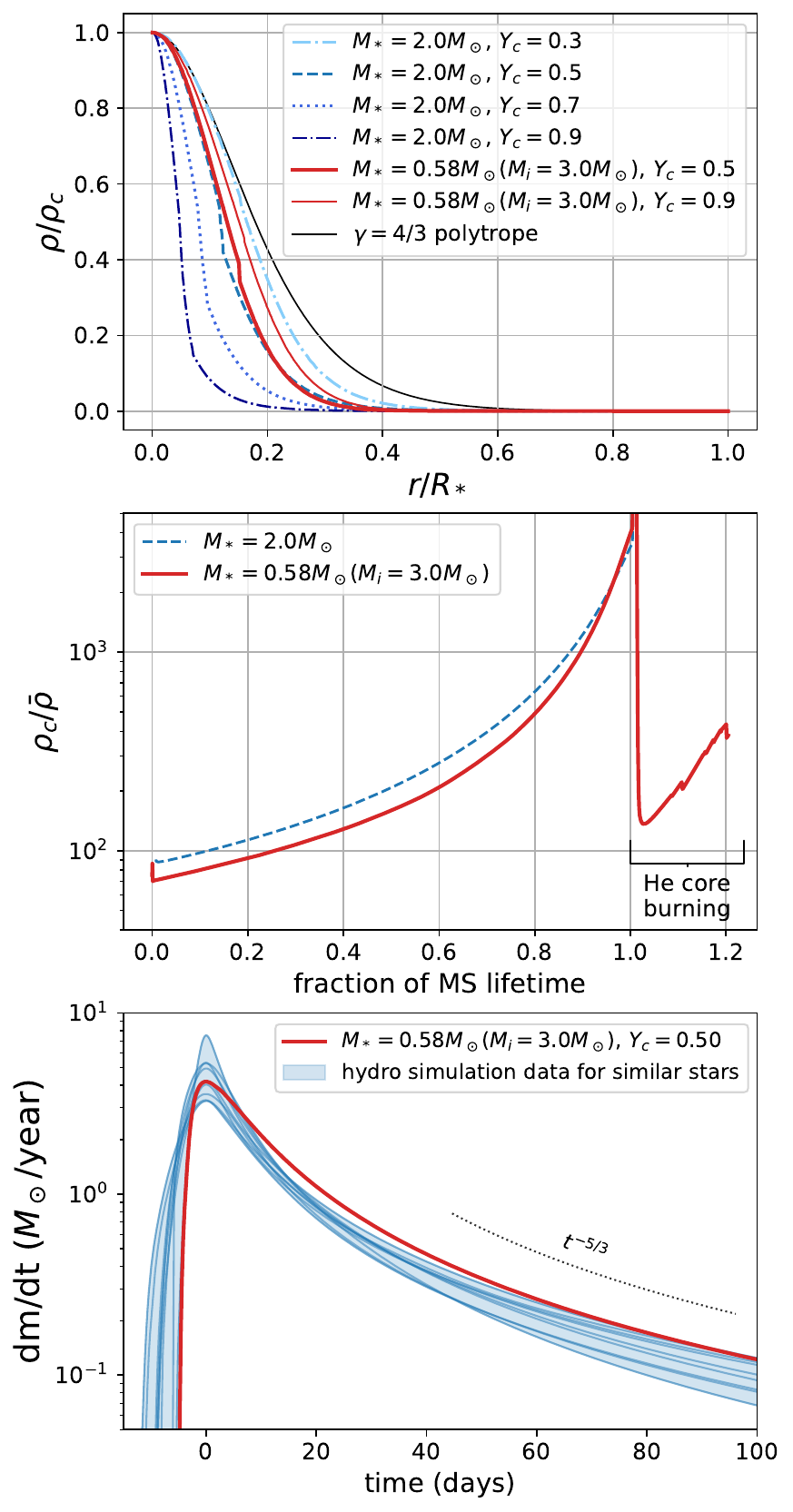}
\end{center}
\vspace{-0.7cm}
\caption{{\bf Top:} Comparison of the density profile of a stripped star ($M_* = 0.58 M_\odot$, $M_i = 3 M_\odot$) with the density profiles of a $2 M_\odot$ star at various stages of stellar evolution. A polytropic stellar profile ($\gamma = 4/3$) is also plotted for comparison. We scale the density profile by the central density ($\rho_c$), and scale the radii by the stellar radius ($R_*$). {\bf Middle:} The central density ($\rho_c$) divided by the average density ($\bar{\rho} = M_*/(4/3 \pi R_*^3)$) for the same stellar models. For the stripped star, we plot $\rho_c/\bar{\rho}$ for both the main sequence (pre-stripping) and the He-core burning sequence (post-stripping). {\bf Bottom:} Mass fallback rates from various hydrodynamical simulations \citep[blue, from][]{law-smith_stellar_2020} compared to the analytical fallback rate calculated for the stripped star (red). The hydro fallback rates are from full disruptions of stars with similar density profiles and central concentrations to the stripped star, scaled to the mass and radius of the stripped star (and to $M_h = 10^{6.8} M_\odot$, a reasonable estimate for ASASSN-14li).  
\label{fig:rho_profile}
}
\end{figure}

The fallback rate of gas to a supermassive black hole after disruption has been well-studied and predictions from hydrodynamic simulations have largely converged \citep[e.g.][]{ramirez-ruiz_star_2009, guillochon_hydrodynamical_2013, cheng_tidal_2014, tejeda_tidal_2017, gafton_tidal_2019, law-smith_stellar_2020, ryu_tidal_2020}. However, previous work focuses on the disruptions of non-interacting stars that evolved through single stellar evolution, and most previous work also focuses on main-sequence stars\footnote{Notable exceptions being papers on red giant disruptions \citep[e.g.][]{macleod_tidal_2012,bogdanovic_disruption_2014}
and white dwarf disruptions \citep[e.g.][]{rosswog_tidal_2009, haas_tidal_2012, law-smith_low-mass_2017, kawana_tidal_2018}.}. 

Stripped stars are more compact (have smaller radii) than main-sequence stars, however we find that their scaled density profiles are actually well approximated by 2 solar mass main sequence stars. Relatedly, their central density divided by their average density ($\rho_c/ \bar{\rho}$) -- a proxy for the disruptability of a star \citep[][]{law-smith_stellar_2020, Ryu:2020a}, is also similar to the value for main-sequence stars. This is convenient, as the density profile and central concentration of a star determines both the critical impact parameter for full disruption \citep[the minimum impact parameter for full disruption, above which all disruptions are full disruptions][]{guillochon_hydrodynamical_2013, law-smith_stellar_2020, ryu_measuring_2020, coughlin_simple_2022}, and the shape of the resulting mass fallback rate to the black hole. 

In Figure~\ref{fig:rho_profile}, we compare the density profiles of our stripped stellar models (taken from \citealt{gotberg_spectral_2018}) with the density profiles of various main sequence stars as well as a $\gamma = 4/3$ polytropic model. We see that the density profile of a $2 M_\odot$ main sequence star with $Y_c = 0.5$ is a very good approximation of our fiducial stripped star model (also with $Y_c = 0.5$). From looking at the evolution between the $Y_c = 0.5$ and $Y_c = 0.9$ models for the MS star and stripped star respectively, it is also clear that the structure of the stripped star changes less dramatically over its Helium burning lifetime as the structure of the MS star changes over its Hydrogen burning lifetime. We also compare the central density over the average density ($\rho_c/\bar{\rho}$) for the $2 M_\odot$ star and the stripped star over their full evolution (main sequence for the $2 M_\odot$ star, main sequence and Helium core burning for the stripped star). The values for $\rho_c/\bar{\rho}$ during Helium core burning for the stripped star are similar to the values for the main sequence star during the middle third of its main sequence evolution. Using the analytical formula for impact parameter as a function of $\rho_c/\bar{\rho}$ from \citet[]{law-smith_stellar_2020} ($\beta_{\rm crit} \approx 0.5 (\rho_c/\bar{\rho})^{1/3}$), we find that this corresponds to critical impact parameters for full disruption between $2.5< \beta_{\rm crit} < 3.5$.

Using the information from the top two panels of Figure~\ref{fig:rho_profile}, in the bottom panel we have plotted a range of hydrodynamical fallback rates from \citet[]{law-smith_stellar_2020} for stars with similar stellar structures to our stripped star\footnote{We included full disruptions of all stars with $36 < \rho_c/\bar{\rho} < 756$ from \citet[]{law-smith_stellar_2020}. These include: TAMS $0.7 M_\odot$ star, ZAMS-TAMS $1 M_\odot$ star, ZAMS $1.5 M_\odot$, ZAMS $3 M_\odot$ star.}. To compare with our analytical fallback rate, we scale these hydrodynamical fallback rates by the mass and radius of the stripped star using the canonical $t_{\rm peak} \propto M_*^{-1}R_*^{3/2}$ and $\dot{M}_{\rm peak}\propto M_*^{2}R_*^{-3/2}$ relations. Given that we have specifically chosen stars with similar density profiles and values of $\rho_c/\bar{\rho}$, this should give consistent peak timescales and fallback rates to the updated equations for these parameters presented in \citet{coughlin_simple_2022} and \citet{bandopadhyay_peak_2023} (which reduce to the canonical relations when holding $\rho_c/\bar{\rho}$ constant).

\subsection{Outflows}\label{sec:outflows} 

Observations of tidal disruption events provide estimates for the size scales of gas around the black hole from the disrupted star through measurements of blackbody photosphere radii and wind radii and velocities \citep[e.g.][]{Hung:2017a, alexander_discovery_2016, Kara:2018, mockler_weighing_2019, wevers_delayed_2023}. Additionally, broad, blue-shifted lines have been observed in many TDEs, consistent with lines forming in dense, outflowing gas \citep[e.g.][]{Arcavi:2014a,Holoien:2016a, nicholl_outflow_2020, Roth:2018}. Lines are expected to form somewhere between the photosphere and outer wind radius, and therefore should be dependent on the composition of the photosphere and outflows. 

We predict that outflows from TDEs of stripped stars will have a Nitrogen enhanced outer layer, with N/C abundance ratios up to $50\times$ higher than is achievable with main-sequence stars (see Figure~\ref{fig:N14_C12}). Emission lines produced by the irradiated outflow would therefore show strong N/C enhancements at early times. Then, as the wind expands and material that returned to the black hole at later times is illuminated, the relative strength of N/C would flip, and Carbon would be enhanced relative to Nitrogen. The exact timing of this behavior and the amount of gas with very high N/C abundance ratios depends on how much mixing occurs. If the gas in the outermost layers of the star that is most bound to the black hole is mixed with material deeper in the star either in the disruption process or in the circularization and accretion processes that launch the outflows, then the N/C abundance will be lower, and will decrease more rapidly.

While hydro simulations are necessary to truly determine the extent of the mixing, we provide some limits here. If the material in the outer layer of the star experiences minimal mixing during the disruption process and outflows are launched promptly after material first returns to the black hole, then it is possible that all material launched in outflows during the rise of the light curve is highly Nitrogen enriched. While there is debate on the length of time between when material begins to return to the black hole and when a flare is produced, most previous results agree that once a bright flare is produced, outflows should also be launched. This could either be through shocks and stream collisions, or through disk winds, or through both processes \citep[e.g.,][]{Dai:2018, bonnerot_first_2021, steinberg_origins_2022, huang_bright_2023, price_eddington_2024}.

In Figure~\ref{fig:outflow_geometry}, we plot the range of radii for a wind launched between model estimates of when material began to return to the black hole and when the light curve peaked for the transient ASASSN-14li, and compare to observations (this event is discussed in more detail in the following section). We assume there is minimal delay between when material returns to the black hole and when luminosity and outflows are produced. If there is a longer delay, material will be N/C enhanced for less time/over a smaller radii range. Regardless, if N/C enhancement is observed from the disruption of a stripped star, the N/C abundance ratio of material surrounding the black hole should decrease with time

\subsection{Comparing to ASASSN-14li}\label{sec:14li}

Nitrogen lines are measured in optical spectra of ASASSN-14li from MJD 56993-57129 \citep{Cenko:2016, Holoien:2016a, Brown:2018}, however the more constraining UV and X-ray detections of high Nitrogen/Carbon abundance ratios occurred between MJD 56999-57002 \citep[XMM and Chandra X-ray spectra,][]{Miller:2015a, miller_evidence_2023} and on MJD 57032 \citep[HST UV spectra,][]{Cenko:2016}. The X-ray spectra constrain the N/C abundance ratio to be $\geq 300$ relative to the sun in material close to the black hole near the peak of the light curve, and the UV spectra constrain the N/C ratio to be $\geq 10$ in material at much larger radii on the decline of the light curve. Interestingly, the N/C ratio in the envelope of the stripped stars we analyze is also $\geq 300$ (see Figure~\ref{fig:N14_C12}). As we show in this section, the measurements from both X-ray and UV observations are consistent with coming from material that returned to the black hole before the peak of the fallback rate, and was then subsequently fed to the accretion disk or ejected in outflows.

\begin{figure}[ht!]
\begin{center}
\includegraphics[width = \columnwidth]{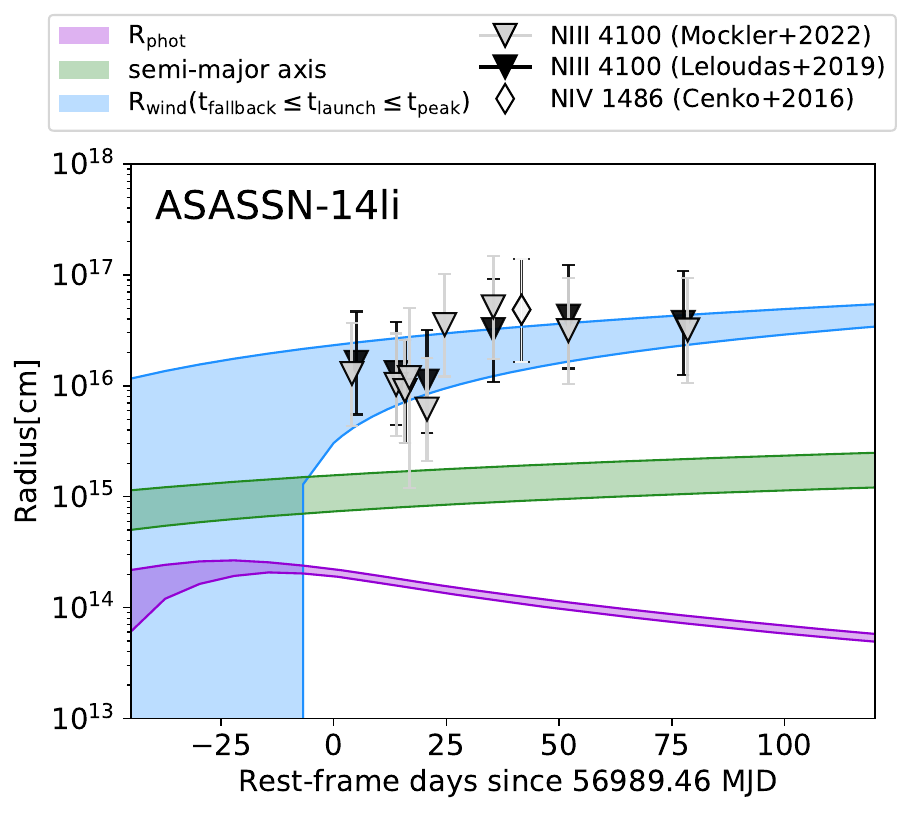}
\end{center}
\vspace{-0.5cm}
\caption{Radii of gas in TDE ASSASN-14li, originally estimated in \citet{mockler_evidence_2022} using the {\tt MOSFiT} transient fitting code. The x-axis is time in rest-frame days since the first observation. We have plotted in blue the radii outflows would reach if launched between first fallback and peak of the mass fallback rate (assuming a velocity of 0.1c, based on radio observations of the event). This corresponds to the time range in the mass fallback rate during which our analytical model predicts a large N/C enhancement. We overplot the virial radii of various Nitrogen lines observed in the event, including a UV measurement (white diamond) that puts constraints on the N/C ratio of the debris. We also plot the radii of the blackbody photosphere and semi-major axis of returning debris in purple and green respectively. 
\label{fig:outflow_geometry}
}
\end{figure}

As discussed in the previous section, if the flare was from the disruption of a stripped star, material that returns to the black hole at early times and up to the peak of the fallback rate \citep[which we are assuming here to be approximately coincident with the peak of the optical and UV light curve, as is predicted by many models, e.g.][]{mockler_weighing_2019, ryu_measuring_2020, steinberg_origins_2022} is likely to be heavily enhanced in Nitrogen with respect to Carbon (see Figure~\ref{fig:N14_C12}). Unfortunately the peak of the transient ASASSN-14li was not observed, however radio observations place rough estimates on the time that outflows were first launched to be between MJD $\sim 56884-56990$ \citep{alexander_discovery_2016}, and light curve analysis place estimates on the peak of the light curve to be between MJD $\sim 56944-56990$ \citep[e.g.,][]{mockler_weighing_2019}. Given this information, the early outflows launched before  MJD $\sim 56990$ and material accreted around the same time would likely have high N/C abundance ratios if the material came from a stripped star. 

There is much theoretical debate on the nature of the initial luminosity production on the rise of TDEs, however multiple theoretical models have shown that it is reasonable for material to start to circularize into a disk by the time of the fallback rate and light curve's peak \citep[e.g.][]{bonnerot_first_2021, steinberg_origins_2022, andalman_tidal_2022}. The observations of X-ray spectra in \citet{miller_evidence_2023} were taken starting on MJD 56999. The X-rays are therefore consistent with being emitted from a disk made up of material that mostly fell back before peak but only recently circularized (and would be Nitrogen-enriched if from a stripped star), combined with additional material that has fallen back after peak. While it is expected that the X-ray emission originates from close to the black hole in the forming disk, the UV emission is consistent with originating from optically thick outflows at larger radii. These outflows would come from gas that returned to the black hole at earlier times, and was then ejected through, e.g.\ stream collisions or disk winds \citep[]{Dai:2018, bonnerot_first_2021}{}. 

In Figure~\ref{fig:outflow_geometry} we show that the virial velocities measured from the widths of Nitrogen lines in the optical and UV spectra are consistent with the range of radii where gas from outflows launched before peak would be at the time of the observations. We plot an estimate of the gas geometry surrounding ASASSN-14li based on light curve fits and the wind velocity measured in radio observations \citep[adapted from][]{mockler_evidence_2022}. If the material is not virialized, and gas broadening dominates the linewidths instead \citep{Roth:2018, parkinson_optical_2022}, the gas could also originate at smaller radii where it would be more likely to be mixed with material from the core of the star that returns at later times. For a stripped star, this would mean material that is not enhanced in N/C (e.g., see Figure~\ref{fig:N14_C12}). The UV spectra constrain the N/C abundance ratio to be $\geq 10 \times$ solar metallicity -- a limit that is much lower than the N/C ratio in the envelope of a stripped star ($N/C \geq 300$). Therefore, even if the lines originate at smaller radii from a mix of gas from the outer layers and the core, this could still be consistent with the disruption of a stripped star.

\section{Discussion}\label{sec:discussion}

Here we discuss additional relevant points that can be investigated further in the future.

\textit{Partial and Repeated TDEs:} If a stripped star is partially disrupted instead of fully disrupted, the core will remain intact and the material that is bound to the black hole will come entirely from the envelope. In this case, all of the material that returns to the black hole will have high N/C abundance ratios (both before and after peak). Partial disruptions are likely intrinsically more common than full disruptions, as two-body relaxation processes produce significantly more partial disruptions than full disruptions \citep[e.g.][]{stone_rates_2016}. However, they are expected to be more difficult to observe. First of all, less accreted material likely means less luminous events. Additionally, if optical and UV emission is due to reprocessing, then less accreted material also means that optical and UV emission will make up a smaller fraction of the total emitted luminosity \citep[e.g.][]{roth_x-ray_2016}. Given this, we expect partial TDEs from stripped stars to produce lower luminosity, X-ray bright TDEs with consistently high N/C ratios.

In the case of {\it repeated} partial disruptions (`repeated TDEs'), the flares should eventually switch from Nitrogen-rich to Nitrogen-poor as subsequent disruptions remove material from deeper within the star. Interestingly, repeated TDEs from stripped stars are theoretically predicted to survive longer and therefore potentially produce more flares than repeated TDEs from main sequence stars. This is because the core fraction is larger in stripped stars than most main sequence stars, which results in the star contracting instead of expanding when a small amount of mass is removed \citep{liu_tidal_2023}. It is possible a stripped star could explain the repeated partial disruption ASASSN-14ko \citep[]{payne_asassn-14ko_2021}, as this event has survived for years without significant evolution observed in its flares \citep[as would be expected if its core mass fraction was $\lesssim 0.3$][]{liu_tidal_2023}. This event actually might be Nitrogen-poor at later times given the appearance of strong CIII lines in the UV without obvious NIII counterparts in the HST spectra taken during flares 18 and 19 \citep[][unfortunately there is no UV spectra from earlier flares to compare with]{payne_chandra_2023}. However, this event also appears to require a star of at least $\sim 2 M_\odot$ to explain the observed luminosity with reasonable mass-to-energy efficiencies \citep[e.g. $\epsilon \sim 0.01$, consistent with predictions of inefficient accretion for these events][]{liu_tidal_2023}, and so a stripped star from a larger mass progenitor ($\gtrsim 7 M_\odot$) would be required \citep{gotberg_spectral_2018}.

\textit{Rates of stripped star TDEs:} A detailed rates calculation is outside the scope of this paper, but a brief overview of how the number of stripped stars compares to the total stellar population is still useful. The fraction of roughly solar mass, solar metallicity stars born in close binaries is $\approx 0.2$ \citep[][]{moe_close_2019}. This fraction increases with stellar mass and decreases with metallicity \citep[e.g. it is $\approx 0.3$ for higher mass stars,][]{geller_wiyn_2012, badenes_stellar_2018, lee_origin_2020, jadhav_high_2021}. It also increases towards the centers of clusters due to mass-segregation \citep[e.g.,][]{mathieu_spatial_1986,geller_wiyn_2012, milone_acs_2012}. Envelope-stripping is completed after the main sequence, meaning that stripped stars can be quite long-lived, as the Helium-core burning stage is left. For massive stars, the stripped star stage correspond to about 10\% of the total stellar lifetime. However, for stars with initial masses $\lesssim 4 M_{\odot}$, the lack of a thick Hydrogen-burning shell stops the Helium core from substantial growth and lower mass stars can therefore live much longer as stripped -- up to half of the total stellar lifetime \citep[e.g.][]{gotberg_spectral_2018, arancibia-rojas_mass_2023}. Given this, we might expect at least $\sim 3-10\%$ of stars in galactic nuclei that are enriched from the CNO cycle to be stripped stars ($0.2 \times 0.5 = 0.1$ for the smallest stars, $0.3 \times 0.1 = 0.03$ for the largest stars). 

The smaller radii and greater compactness of stripped stars make them more difficult to partially disrupt than main-sequence stars, however, their critical disruption radii for core disruption are less than a factor of 10 different\footnote{For example, the critical disruption radius for our fiducial stripped star is within a factor of 5-6 of the critical disruption radius for the $2 M_\odot$ main-sequence star we compare it to. See Section~\ref{sec:fallback} for a more in depth discussion of how the critical disruption radius is determined.} than the critical radii for main-sequence stars (as might be expected given their similar values of $\rho_c/\bar{\rho}$). 
Given that the TDE rate from two-body relaxation scales as $\propto R_t^{1/4}$ for a wide range of tidal radii and black hole masses \citep[e.g.][]{wang_revised_2004, milosavljevic_contribution_2006, MacLeod:2012a}, the likelihood of full disruption for stars after being stripped and during their Helium burning sequence is, at any given time, within a factor of two ($\lesssim 10^{1/4} = 1.78$) of the likelihood of full disruption for the same stars during their main sequence. Therefore we might expect at least $\sim 1 - 5 \%$ of the disruptions of stars {\it born} at $M_* \gtrsim 1 M_\odot$ to come from stripped stars (including other factors such as the age and mass-segregation of stellar populations should increase this percentage). We can organize this estimate as follows:

\begin{equation}
    f_{\rm TDE, \: stripped} \approx f_{\rm close \: binary} \times \frac{\tau_{\rm stripped}}{\tau_{\rm *}} \times \Big(\frac{R_{t,\: \rm stripped}}{R_{t,\: \rm MS}}\Big)^{1/4}
\end{equation}

Here $f_{\rm close \: binary}$ is the fraction of stars born in closed binaries that are expected to end their lives as stripped stars -- $f_{\rm close \: binary}\sim 0.2 (0.3)$ for low mass (high mass) stars. Then $\tau_{\rm stripped}$ and $\tau_{*}$ are the stripped star and total stellar (MS $+$ stripped) lifetimes -- $\frac{\tau_{\rm stripped}}{\tau_{\rm *}} \sim 0.5 (0.1)$ for low mass (high mass) stars. And finally, $R_{t,\: \rm stripped}$ and $R_{t,\: \rm MS}$ are the tidal radii of stripped stars and MS stars respectively -- $\frac{R_{t,\: \rm stripped}}{R_{t,\: \rm MS}} \gtrsim 0.1$ for the majority of binary stripped stars (using stripped star parameters from \citealt{gotberg_spectral_2018} throughout). These leads to $f_{\rm TDE, \: stripped} \sim 0.056 (0.017)$ for low mass (high mass) stars, or $\sim 1-5\%$ of TDEs from stars born at masses such that they will undergo CNO burning ($M_* \gtrsim 1 M_\odot$).

\textit{Connections to Nitrogen-enrichment in AGN disks:}

Regions associated with galactic centers have been found to be Nitrogen enriched, for example through the spectra of AGN broad line regions (BLR) \citep[e.g.][]{bentz_nitrogen-enriched_2004,  batra_metallicities_2014}. TDE searches have, up to this point, mostly excluded AGN host galaxies, and so a pre-existing Nitrogen-rich BLR is disfavored to explain Nitrogen-rich TDEs. However, it is still possible that AGN and TDEs share the same source of Nitrogen-enrichment.

The centers of galaxies also often have higher than solar overall metallicities. Some studies have found that all BLR line measurements in those galaxies (not just N/C ratios) are consistent with super-solar metallicities \citep[e.g.][]{batra_metallicities_2014}, and therefore Nitrogen enhancements can be explained by these higher overall metallicities. Others find that the enhancement of Nitrogen is higher than that of other metals \citep{jiang_sample_2008, matsuoka_chemical_2017}, and therefore can't be explained solely by increasing the overall metallicity of the galactic nuclei. The most common explanation put forward for general metallicity enhancement is high star formation rates in the vicinity of the SMBH (consistent with observations of star formation gradients in galaxies \citealt[e.g.][]{nelson_where_2016}, but at size scales not easily testable observationally). To explain anomalously high Nitrogen abundances with respect to the overall metallicity, \citet{matsuoka_chemical_2017} invoke mass loss from AGB winds enriching the BLR gas with Nitrogen (which was dredged up from CNO processed material near the core). This scenario also requires in-situ star formation to create sufficient AGB stars. Additionally, it predicts higher AGN accretion rates as some of the winds will end up accreted by the SMBH \citep{davies_close_2007}. It does appear that Nitrogen-loud quasars have higher Eddington ratios than the general population, in line with the prediction of AGB winds enriching BLR clouds and also increasing SMBH accretion rates \citep{matsuoka_chemical_2017}.

High star formation rates in the centers of galaxies would also help explain the Nitrogen enhancement found in some TDEs, as it would increase the fraction of higher mass stars undergoing CNO cycle burning. However, high star formation rates alone are likely not sufficient to explain the unexpectedly high rate of Nitrogen enhanced TDEs \citep{mockler_evidence_2022}. Additionally, higher overall metallicities in the centers of galaxies are not sufficient to explain the particularly extreme Nitrogen enhancement found in ASASSN-14li (see Figures~\ref{fig:dmdt_composition},~\ref{fig:N14_C12}), unless these high birth metallicities are accompanied by unusually high N/C abundance ratios, or young stars are polluted by winds from massive stars. These two scenarios have been suggested to explain highly Nitrogen-enhanced stars in globular clusters \citep[although these globular cluster stars are still less Nitrogen-enhanced than ASASSN-14li appears to be][]{larsen_nitrogen_2014,bastian_multiple_2018}. On the other hand, TDEs can clearly contribute to the Nitrogen-enhancement of a SMBH's environment, and it is possible that this contribution is important to the population of Nitrogen-enhanced AGN. This connection is particularly intriguing if the galaxy is in a post-starburst phase of evolution, or if the nuclear cluster hosts a SMBH binary given that these environments appear 1) conducive to AGN activity, \citep[e.g.][]{dodd_landscape_2021}, and 2) produce TDE rates that are observed or predicted to be $\gtrsim 20-100 \times$ higher than average, \citep[e.g.][]{french_tidal_2016, graur_dependence_2018, law-smith_tidal_2019, french_host_2020, hammerstein_tidal_2021, melchor_tidal_2023-1, mockler_uncovering_2023}.

\section{Summary of Key Points}
Here we briefly summarize the main takeaways from this work: 
\begin{itemize}
    \item TDEs probe stellar populations in galactic nuclei, and recent observations of N/C abundances in TDEs point to the disruption of stars born at moderately high masses that show chemical enrichment indicative of the CNO cycle.
    
    \item X-ray spectral observations of ASASSN-14li show N/C abundance lower limits so high ($\gtrsim 300$) it is difficult to explain with main-sequence evolution alone \citep{miller_evidence_2023}. In a main-sequence star, the core has high N/C ratios but the envelope is more Carbon than Nitrogen. In a TDE, when material from the core eventually falls back to the black hole it is diluted by the Carbon in the envelope material that falls back with it, limiting the maximum N/C abundance in fallback material (see Figure~\ref{fig:N14_C12}).
    
    \item Binary-stripped stars have the reverse compositional structure of main-sequence stars. Their Helium-rich envelopes have high N/C abundance ratios, and it is their cores that are Nitrogen-poor and Carbon-enhanced (see Figure~\ref{fig:composition_mass_fraction}). When these stars are tidally disrupted, Nitrogen-rich material from the envelope will return to the black hole first, without being diluted by Nitrogen-poor material from the core. This means the composition of fallback material from a tidally disrupted stripped star can have much higher enhancements of N/C than disruptions of main-sequence stars.
    
    \item Stripped stars also naturally produce the opposite N/C time evolution as main-sequence stars and should therefore be observationally distinct in time series of their spectra (see Figure~\ref{fig:N14_C12}).
\end{itemize}

\section*{Acknowledgments}
We thank the participants and organizers of the summer Aspen 2023 workshop on ``Stellar Interactions and the Transients They Cause" for fruitful discussions. 
B.M. is grateful for support from the Carnegie Theoretical Astrophysics Center. M.G.G. is grateful for the support from Northwestern University's Presidential Fellowship. E.R.R. thank the Heising-Simons Foundation, NSF (AST-2150255 and AST-2307710), Swift (80NSSC21K1409, 80NSSC19K1391) and Chandra (22-0142) for support. 

\vspace{5mm}

\software{astropy \citep{Astropy-Collaboration:2013a} 
          }



\bibliography{zotero_library, library}{}

\begin{thebibliography}{}
\expandafter\ifx\csname natexlab\endcsname\relax\def\natexlab#1{#1}\fi
\providecommand{\url}[1]{\href{#1}{#1}}
\providecommand{\dodoi}[1]{doi:~\href{http://doi.org/#1}{\nolinkurl{#1}}}
\providecommand{\doeprint}[1]{\href{http://ascl.net/#1}{\nolinkurl{http://ascl.net/#1}}}
\providecommand{\doarXiv}[1]{\href{https://arxiv.org/abs/#1}{\nolinkurl{https://arxiv.org/abs/#1}}}

\bibitem[{Alexander {et~al.}(2016)Alexander, Berger, Guillochon, Zauderer, \&
  Williams}]{alexander_discovery_2016}
Alexander, K.~D., Berger, E., Guillochon, J., Zauderer, B.~A., \& Williams, P.
  K.~G. 2016, The Astrophysical Journal, 819, L25,
  \dodoi{10.3847/2041-8205/819/2/L25}

\bibitem[{Andalman {et~al.}(2022)Andalman, Liska, Tchekhovskoy, Coughlin, \&
  Stone}]{andalman_tidal_2022}
Andalman, Z.~L., Liska, M. T.~P., Tchekhovskoy, A., Coughlin, E.~R., \& Stone,
  N. 2022, Monthly Notices of the Royal Astronomical Society, 510, 1627,
  \dodoi{10.1093/mnras/stab3444}

\bibitem[{Arancibia-Rojas {et~al.}(2023)Arancibia-Rojas, Zorotovic, Vučković,
  Bobrick, Vos, \& Piraino-Cerda}]{arancibia-rojas_mass_2023}
Arancibia-Rojas, E., Zorotovic, M., Vučković, M., {et~al.} 2023, Monthly
  Notices of the Royal Astronomical Society, \dodoi{10.1093/mnras/stad3891}

\bibitem[{{Arcavi} {et~al.}(2014){Arcavi}, {Gal-Yam}, {Sullivan}, {Pan},
  {Cenko}, {Horesh}, {Ofek}, {De Cia}, {Yan}, {Yang}, {Howell}, {Tal},
  {Kulkarni}, {Tendulkar}, {Tang}, {Xu}, {Sternberg}, {Cohen}, {Bloom},
  {Nugent}, {Kasliwal}, {Perley}, {Quimby}, {Miller}, {Theissen}, \&
  {Laher}}]{Arcavi:2014a}
{Arcavi}, I., {Gal-Yam}, A., {Sullivan}, M., {et~al.} 2014, \apj, 793, 38,
  \dodoi{10.1088/0004-637X/793/1/38}

\bibitem[{{Astropy Collaboration} {et~al.}(2013){Astropy Collaboration},
  {Robitaille}, {Tollerud}, {Greenfield}, {Droettboom}, {Bray}, {Aldcroft},
  {Davis}, {Ginsburg}, {Price-Whelan}, {Kerzendorf}, {Conley}, {Crighton},
  {Barbary}, {Muna}, {Ferguson}, {Grollier}, {Parikh}, {Nair}, {Unther},
  {Deil}, {Woillez}, {Conseil}, {Kramer}, {Turner}, {Singer}, {Fox}, {Weaver},
  {Zabalza}, {Edwards}, {Azalee Bostroem}, {Burke}, {Casey}, {Crawford},
  {Dencheva}, {Ely}, {Jenness}, {Labrie}, {Lim}, {Pierfederici}, {Pontzen},
  {Ptak}, {Refsdal}, {Servillat}, \& {Streicher}}]{Astropy-Collaboration:2013a}
{Astropy Collaboration}, {Robitaille}, T.~P., {Tollerud}, E.~J., {et~al.} 2013,
  \aap, 558, A33, \dodoi{10.1051/0004-6361/201322068}

\bibitem[{Badenes {et~al.}(2018)Badenes, Mazzola, Thompson, Covey, Freeman,
  Walker, Moe, Troup, Nidever, Allende~Prieto, Andrews, Barbá, Beers, Bovy,
  Carlberg, De~Lee, Johnson, Lewis, Majewski, Pinsonneault, Sobeck, Stassun,
  Stringfellow, \& Zasowski}]{badenes_stellar_2018}
Badenes, C., Mazzola, C., Thompson, T.~A., {et~al.} 2018, The Astrophysical
  Journal, 854, 147, \dodoi{10.3847/1538-4357/aaa765}

\bibitem[{Bandopadhyay {et~al.}(2023)Bandopadhyay, Fancher, Athian, Indelicato,
  Kapalanga, Kumah, Paradiso, Todd, Coughlin, \&
  Nixon}]{bandopadhyay_peak_2023}
Bandopadhyay, A., Fancher, J., Athian, A., {et~al.} 2023, The {Peak} of the
  {Fallback} {Rate} from {Tidal} {Disruption} {Events}: {Dependence} on
  {Stellar} {Type}, \dodoi{10.48550/arXiv.2310.11496}

\bibitem[{Bastian \& Lardo(2018)}]{bastian_multiple_2018}
Bastian, N., \& Lardo, C. 2018, Annual Review of Astronomy and Astrophysics,
  56, 83, \dodoi{10.1146/annurev-astro-081817-051839}

\bibitem[{Batra \& Baldwin(2014)}]{batra_metallicities_2014}
Batra, N.~D., \& Baldwin, J.~A. 2014, Monthly Notices of the Royal Astronomical
  Society, 439, 771, \dodoi{10.1093/mnras/stu007}

\bibitem[{Bentz {et~al.}(2004)Bentz, Hall, \&
  Osmer}]{bentz_nitrogen-enriched_2004}
Bentz, M.~C., Hall, P.~B., \& Osmer, P.~S. 2004, The Astronomical Journal, 128,
  561, \dodoi{10.1086/422346}

\bibitem[{Blagorodnova {et~al.}(2017)Blagorodnova, Gezari, Hung, Kulkarni,
  Cenko, Pasham, Yan, Arcavi, Ben-Ami, Bue, Cantwell, Cao, Castro-Tirado,
  Fender, Fremling, Gal-Yam, Ho, Horesh, Hosseinzadeh, Kasliwal, Kong, Laher,
  Leloudas, Lunnan, Masci, Mooley, Neill, Nugent, Powell, Valeev, Vreeswijk,
  Walters, \& Wozniak}]{blagorodnova_iptf16fnl_2017}
Blagorodnova, N., Gezari, S., Hung, T., {et~al.} 2017, The Astrophysical
  Journal, 844, 46, \dodoi{10.3847/1538-4357/aa7579}

\bibitem[{Blagorodnova {et~al.}(2019)Blagorodnova, Cenko, Kulkarni, Arcavi,
  Bloom, Duggan, Filippenko, Fremling, Horesh, Hosseinzadeh, Karamehmetoglu,
  Levan, Masci, Nugent, Pasham, Veilleux, Walters, Yan, \&
  Zheng}]{blagorodnova_broad_2019}
Blagorodnova, N., Cenko, S.~B., Kulkarni, S.~R., {et~al.} 2019, The
  Astrophysical Journal, 873, 92, \dodoi{10.3847/1538-4357/ab04b0}

\bibitem[{Bogdanović {et~al.}(2014)Bogdanović, Cheng, \&
  Amaro-Seoane}]{bogdanovic_disruption_2014}
Bogdanović, T., Cheng, R.~M., \& Amaro-Seoane, P. 2014, The Astrophysical
  Journal, 788, 99, \dodoi{10.1088/0004-637X/788/2/99}

\bibitem[{Bonnerot {et~al.}(2021)Bonnerot, Lu, \&
  Hopkins}]{bonnerot_first_2021}
Bonnerot, C., Lu, W., \& Hopkins, P.~F. 2021, Monthly Notices of the Royal
  Astronomical Society, 504, 4885, \dodoi{10.1093/mnras/stab398}

\bibitem[{{Bricman} \& {Gomboc}(2020)}]{Bricman:2020}
{Bricman}, K., \& {Gomboc}, A. 2020, \apj, 890, 73,
  \dodoi{10.3847/1538-4357/ab6989}

\bibitem[{{Brown} {et~al.}(2018){Brown}, {Kochanek}, {Holoien}, {Stanek},
  {Auchettl}, {Shappee}, {Prieto}, {Morrell}, {Falco}, {Strader}, {Chomiuk},
  {Post}, {Villanueva}, {Mathur}, {Dong}, {Chen}, \& {Bose}}]{Brown:2018}
{Brown}, J.~S., {Kochanek}, C.~S., {Holoien}, T.~W.~S., {et~al.} 2018, \mnras,
  473, 1130, \dodoi{10.1093/mnras/stx2372}

\bibitem[{Bunker {et~al.}(2023)Bunker, Saxena, Cameron, Willott, Curtis-Lake,
  Jakobsen, Carniani, Smit, Maiolino, Witstok, Curti, D'Eugenio, Jones,
  Ferruit, Arribas, Charlot, Chevallard, Giardino, de~Graaff, Looser,
  Lützgendorf, Maseda, Rawle, Rix, Del~Pino, Alberts, Egami, Eisenstein,
  Endsley, Hainline, Hausen, Johnson, Rieke, Rieke, Robertson, Shivaei, Stark,
  Sun, Tacchella, Tang, Williams, Willmer, Baker, Baum, Bhatawdekar, Bowler,
  Boyett, Chen, Circosta, Helton, Ji, Kumari, Lyu, Nelson, Parlanti, Perna,
  Sandles, Scholtz, Suess, Topping, Übler, Wallace, \&
  Whitler}]{bunker_jades_2023}
Bunker, A.~J., Saxena, A., Cameron, A.~J., {et~al.} 2023, Astronomy and
  Astrophysics, 677, A88, \dodoi{10.1051/0004-6361/202346159}

\bibitem[{Burbidge {et~al.}(1957)Burbidge, Burbidge, Fowler, \&
  Hoyle}]{burbidge_synthesis_1957}
Burbidge, E.~M., Burbidge, G.~R., Fowler, W.~A., \& Hoyle, F. 1957, Reviews of
  Modern Physics, 29, 547, \dodoi{10.1103/RevModPhys.29.547}

\bibitem[{Cenko {et~al.}(2016{\natexlab{a}})Cenko, Cucchiara, Roth, Veilleux,
  Prochaska, Yan, Guillochon, Maksym, Arcavi, Butler, Filippenko, Fruchter,
  Gezari, Kasen, Levan, Miller, Pasham, Ramirez-Ruiz, Strubbe, Tanvir, \&
  Tombesi}]{cenko_ultraviolet_2016}
Cenko, S.~B., Cucchiara, A., Roth, N., {et~al.} 2016{\natexlab{a}}, The
  Astrophysical Journal, 818, L32, \dodoi{10.3847/2041-8205/818/2/L32}

\bibitem[{Cenko {et~al.}(2016{\natexlab{b}})Cenko, Cucchiara, Roth, Veilleux,
  Prochaska, Yan, Guillochon, Maksym, Arcavi, Butler, Filippenko, Fruchter,
  Gezari, Kasen, Levan, Miller, Pasham, Ramirez-Ruiz, Strubbe, Tanvir, \&
  Tombesi}]{Cenko:2016}
---. 2016{\natexlab{b}}, The Astrophysical Journal, 818, L32,
  \dodoi{10.3847/2041-8205/818/2/l32}

\bibitem[{Cheng \& Bogdanović(2014)}]{cheng_tidal_2014}
Cheng, R.~M., \& Bogdanović, T. 2014, Physical Review D, 90, 064020,
  \dodoi{10.1103/PhysRevD.90.064020}

\bibitem[{Coughlin \& Nixon(2022)}]{coughlin_simple_2022}
Coughlin, E.~R., \& Nixon, C.~J. 2022, Monthly Notices of the Royal
  Astronomical Society, 517, L26, \dodoi{10.1093/mnrasl/slac106}

\bibitem[{{Dai} {et~al.}(2018){Dai}, {McKinney}, {Roth}, {Ramirez-Ruiz}, \&
  {Miller}}]{Dai:2018}
{Dai}, L., {McKinney}, J.~C., {Roth}, N., {Ramirez-Ruiz}, E., \& {Miller},
  M.~C. 2018, \apjl, 859, L20, \dodoi{10.3847/2041-8213/aab429}

\bibitem[{Davies {et~al.}(2007)Davies, Müller~Sánchez, Genzel, Tacconi,
  Hicks, Friedrich, \& Sternberg}]{davies_close_2007}
Davies, R.~I., Müller~Sánchez, F., Genzel, R., {et~al.} 2007, The
  Astrophysical Journal, 671, 1388, \dodoi{10.1086/523032}

\bibitem[{Dessart {et~al.}(2012)Dessart, Hillier, Li, \&
  Woosley}]{dessart_nature_2012}
Dessart, L., Hillier, D.~J., Li, C., \& Woosley, S. 2012, Monthly Notices of
  the Royal Astronomical Society, 424, 2139,
  \dodoi{10.1111/j.1365-2966.2012.21374.x}

\bibitem[{Dessart {et~al.}(2011)Dessart, Hillier, Livne, Yoon, Woosley,
  Waldman, \& Langer}]{dessart_core-collapse_2011}
Dessart, L., Hillier, D.~J., Livne, E., {et~al.} 2011, Monthly Notices of the
  Royal Astronomical Society, 414, 2985,
  \dodoi{10.1111/j.1365-2966.2011.18598.x}

\bibitem[{Dodd {et~al.}(2021)Dodd, Law-Smith, Auchettl, Ramirez-Ruiz, \&
  Foley}]{dodd_landscape_2021}
Dodd, S.~A., Law-Smith, J. A.~P., Auchettl, K., Ramirez-Ruiz, E., \& Foley,
  R.~J. 2021, The Astrophysical Journal, 907, L21,
  \dodoi{10.3847/2041-8213/abd852}

\bibitem[{Drilling {et~al.}(2013)Drilling, Jeffery, Heber, Moehler, \&
  Napiwotzki}]{drilling_mk-like_2013}
Drilling, J.~S., Jeffery, C.~S., Heber, U., Moehler, S., \& Napiwotzki, R.
  2013, Astronomy and Astrophysics, 551, A31,
  \dodoi{10.1051/0004-6361/201219433}

\bibitem[{Evans \& Kochanek(1989)}]{Evans:1989a}
Evans, C.~R., \& Kochanek, C.~S. 1989, \apjl, 346, L13.
\newblock
  \url{http://adsabs.harvard.edu/cgi-bin/nph-data_query?bibcode=1989ApJ...346L..13E&link_type=ABSTRACT}

\bibitem[{French {et~al.}(2016)French, Arcavi, \&
  Zabludoff}]{french_tidal_2016}
French, K.~D., Arcavi, I., \& Zabludoff, A. 2016, The Astrophysical Journal,
  818, L21, \dodoi{10.3847/2041-8205/818/1/L21}

\bibitem[{French {et~al.}(2020)French, Wevers, Law-Smith, Graur, \&
  Zabludoff}]{french_host_2020}
French, K.~D., Wevers, T., Law-Smith, J., Graur, O., \& Zabludoff, A.~I. 2020,
  Space Science Reviews, 216, 32, \dodoi{10.1007/s11214-020-00657-y}

\bibitem[{Gafton \& Rosswog(2019)}]{gafton_tidal_2019}
Gafton, E., \& Rosswog, S. 2019, Monthly Notices of the Royal Astronomical
  Society, 487, 4790, \dodoi{10.1093/mnras/stz1530}

\bibitem[{{Gallegos-Garcia} {et~al.}(2018){Gallegos-Garcia}, {Law-Smith}, \&
  {Ramirez-Ruiz}}]{Gallegos-Garcia:2018}
{Gallegos-Garcia}, M., {Law-Smith}, J., \& {Ramirez-Ruiz}, E. 2018, \apj, 857,
  109, \dodoi{10.3847/1538-4357/aab5b8}

\bibitem[{Geller \& Mathieu(2012)}]{geller_wiyn_2012}
Geller, A.~M., \& Mathieu, R.~D. 2012, The Astronomical Journal, 144, 54,
  \dodoi{10.1088/0004-6256/144/2/54}

\bibitem[{Graur {et~al.}(2018)Graur, French, Zahid, Guillochon, Mandel,
  Auchettl, \& Zabludoff}]{graur_dependence_2018}
Graur, O., French, K.~D., Zahid, H.~J., {et~al.} 2018, The Astrophysical
  Journal, 853, 39, \dodoi{10.3847/1538-4357/aaa3fd}

\bibitem[{{Guillochon} \& {Ramirez-Ruiz}(2013)}]{Guillochon:2013a}
{Guillochon}, J., \& {Ramirez-Ruiz}, E. 2013, \apj, 767, 25,
  \dodoi{10.1088/0004-637X/767/1/25}

\bibitem[{Guillochon \& Ramirez-Ruiz(2013)}]{guillochon_hydrodynamical_2013}
Guillochon, J., \& Ramirez-Ruiz, E. 2013, The Astrophysical Journal, 767, 25,
  \dodoi{10.1088/0004-637X/767/1/25}

\bibitem[{Götberg {et~al.}(2018)Götberg, de~Mink, Groh, Kupfer, Crowther,
  Zapartas, \& Renzo}]{gotberg_spectral_2018}
Götberg, Y., de~Mink, S.~E., Groh, J.~H., {et~al.} 2018, Astronomy and
  Astrophysics, 615, A78, \dodoi{10.1051/0004-6361/201732274}

\bibitem[{Götberg {et~al.}(2023)Götberg, Drout, Ji, Groh, Ludwig, Crowther,
  Smith, de~Koter, \& de~Mink}]{gotberg_stellar_2023}
Götberg, Y., Drout, M.~R., Ji, A.~P., {et~al.} 2023, The Astrophysical
  Journal, 959, 125, \dodoi{10.3847/1538-4357/ace5a3}

\bibitem[{Haas {et~al.}(2012)Haas, Shcherbakov, Bode, \&
  Laguna}]{haas_tidal_2012}
Haas, R., Shcherbakov, R.~V., Bode, T., \& Laguna, P. 2012, The Astrophysical
  Journal, 749, 117, \dodoi{10.1088/0004-637X/749/2/117}

\bibitem[{Hammerstein {et~al.}(2021)Hammerstein, Gezari, van Velzen, Cenko,
  Roth, Ward, Frederick, Hung, Graham, Foley, Bellm, Cannella, Drake, Kupfer,
  Laher, Mahabal, Masci, Riddle, Rojas-Bravo, \&
  Smith}]{hammerstein_tidal_2021}
Hammerstein, E., Gezari, S., van Velzen, S., {et~al.} 2021, The Astrophysical
  Journal, 908, L20, \dodoi{10.3847/2041-8213/abdcb4}

\bibitem[{Hammerstein {et~al.}(2023)Hammerstein, van Velzen, Gezari, Cenko,
  Yao, Ward, Frederick, Villanueva, Somalwar, Graham, Kulkarni, Stern,
  Andreoni, Bellm, Dekany, Dhawan, Drake, Fremling, Gatkine, Groom, Ho,
  Kasliwal, Karambelkar, Kool, Masci, Medford, Perley, Purdum, van Roestel,
  Sharma, Sollerman, Taggart, \& Yan}]{hammerstein_final_2023}
Hammerstein, E., van Velzen, S., Gezari, S., {et~al.} 2023, The Astrophysical
  Journal, 942, 9, \dodoi{10.3847/1538-4357/aca283}

\bibitem[{Heber(2016)}]{heber_hot_2016}
Heber, U. 2016, Publications of the Astronomical Society of the Pacific, 128,
  082001, \dodoi{10.1088/1538-3873/128/966/082001}

\bibitem[{Hirsch(2009)}]{hirsch_hot_2009}
Hirsch, H.~A. 2009, PhD thesis.
\newblock \url{https://ui.adsabs.harvard.edu/abs/2009PhDT.......273H}

\bibitem[{{Holoien} {et~al.}(2016){Holoien}, {Kochanek}, {Prieto}, {Stanek},
  {Dong}, {Shappee}, {Grupe}, {Brown}, {Basu}, {Beacom}, {Bersier},
  {Brimacombe}, {Danilet}, {Falco}, {Guo}, {Jose}, {Herczeg}, {Long},
  {Pojmanski}, {Simonian}, {Szczygie{\l}}, {Thompson}, {Thorstensen}, {Wagner},
  \& {Wo{\'z}niak}}]{Holoien:2016a}
{Holoien}, T.~W.~S., {Kochanek}, C.~S., {Prieto}, J.~L., {et~al.} 2016, \mnras,
  455, 2918, \dodoi{10.1093/mnras/stv2486}

\bibitem[{Huang {et~al.}(2023)Huang, Davis, \& Jiang}]{huang_bright_2023}
Huang, X., Davis, S.~W., \& Jiang, Y.-f. 2023, A {Bright} {First} {Day} for
  {Tidal} {Disruption} {Event}, Tech. rep., \dodoi{10.48550/arXiv.2303.17443}

\bibitem[{Hung {et~al.}(2017)Hung, Gezari, Blagorodnova, Roth, Cenko, Kulkarni,
  Horesh, Arcavi, McCully, Yan, Lunnan, Fremling, Cao, Nugent, \&
  Wozniak}]{Hung:2017a}
Hung, T., Gezari, S., Blagorodnova, N., {et~al.} 2017, The Astrophysical
  Journal, 842, 29.
\newblock \url{http://stacks.iop.org/0004-637X/842/i=1/a=29}

\bibitem[{Hung {et~al.}(2019)Hung, Cenko, Roth, Gezari, Veilleux, van Velzen,
  Gaskell, Foley, Blagorodnova, Yan, Graham, Brown, Siebert, Frederick, Ward,
  Gatkine, Gal-Yam, Yang, Schulze, Dimitriadis, Kupfer, Shupe, Rusholme, Masci,
  Riddle, Soumagnac, van Roestel, \& Dekany}]{hung_discovery_2019}
Hung, T., Cenko, S.~B., Roth, N., {et~al.} 2019, The Astrophysical Journal,
  879, 119, \dodoi{10.3847/1538-4357/ab24de}

\bibitem[{Hung {et~al.}(2021)Hung, Foley, Veilleux, Cenko, Dai, Auchettl,
  Brink, Dimitriadis, Filippenko, Gezari, Holoien, Kilpatrick, Mockler, Piro,
  Ramirez-Ruiz, Rojas-Bravo, Siebert, van Velzen, \&
  Zheng}]{hung_discovery_2021}
Hung, T., Foley, R.~J., Veilleux, S., {et~al.} 2021, The Astrophysical Journal,
  917, 9, \dodoi{10.3847/1538-4357/abf4c3}

\bibitem[{Jadhav {et~al.}(2021)Jadhav, Roy, Joshi, \&
  Subramaniam}]{jadhav_high_2021}
Jadhav, V.~V., Roy, K., Joshi, N., \& Subramaniam, A. 2021, The Astronomical
  Journal, 162, 264, \dodoi{10.3847/1538-3881/ac2571}

\bibitem[{Jermyn {et~al.}(2023)Jermyn, Bauer, Schwab, Farmer, Ball, Bellinger,
  Dotter, Joyce, Marchant, Mombarg, Wolf, Sunny~Wong, Cinquegrana, Farrell,
  Smolec, Thoul, Cantiello, Herwig, Toloza, Bildsten, Townsend, \&
  Timmes}]{jermyn_modules_2023}
Jermyn, A.~S., Bauer, E.~B., Schwab, J., {et~al.} 2023, The Astrophysical
  Journal Supplement Series, 265, 15, \dodoi{10.3847/1538-4365/acae8d}

\bibitem[{Jiang {et~al.}(2008)Jiang, Fan, \& Vestergaard}]{jiang_sample_2008}
Jiang, L., Fan, X., \& Vestergaard, M. 2008, The Astrophysical Journal, 679,
  962, \dodoi{10.1086/587868}

\bibitem[{{Kara} {et~al.}(2018){Kara}, {Dai}, {Reynolds}, \&
  {Kallman}}]{Kara:2018}
{Kara}, E., {Dai}, L., {Reynolds}, C.~S., \& {Kallman}, T. 2018, \mnras, 474,
  3593, \dodoi{10.1093/mnras/stx3004}

\bibitem[{Kawana {et~al.}(2018)Kawana, Tanikawa, \&
  Yoshida}]{kawana_tidal_2018}
Kawana, K., Tanikawa, A., \& Yoshida, N. 2018, Monthly Notices of the Royal
  Astronomical Society, 477, 3449, \dodoi{10.1093/mnras/sty842}

\bibitem[{Kesden(2012)}]{Kesden:2012b}
Kesden, M. 2012, \prd, 85, 24037.
\newblock
  \url{http://adsabs.harvard.edu/cgi-bin/nph-data_query?bibcode=2012PhRvD..85b4037K&link_type=ABSTRACT}

\bibitem[{{Kochanek}(2016)}]{Kochanek:2016a}
{Kochanek}, C.~S. 2016, \mnras, 461, 371, \dodoi{10.1093/mnras/stw1290}

\bibitem[{Laplace {et~al.}(2021)Laplace, Justham, Renzo, Götberg, Farmer,
  Vartanyan, \& de~Mink}]{laplace_different_2021}
Laplace, E., Justham, S., Renzo, M., {et~al.} 2021, Astronomy and Astrophysics,
  656, A58, \dodoi{10.1051/0004-6361/202140506}

\bibitem[{Larsen {et~al.}(2014)Larsen, Brodie, Grundahl, \&
  Strader}]{larsen_nitrogen_2014}
Larsen, S.~S., Brodie, J.~P., Grundahl, F., \& Strader, J. 2014, The
  Astrophysical Journal, 797, 15, \dodoi{10.1088/0004-637X/797/1/15}

\bibitem[{Law-Smith {et~al.}(2019)Law-Smith, Guillochon, \&
  Ramirez-Ruiz}]{law-smith_tidal_2019}
Law-Smith, J., Guillochon, J., \& Ramirez-Ruiz, E. 2019, The Astrophysical
  Journal, 882, L25, \dodoi{10.3847/2041-8213/ab379a}

\bibitem[{Law-Smith {et~al.}(2017)Law-Smith, MacLeod, Guillochon, Macias, \&
  Ramirez-Ruiz}]{law-smith_low-mass_2017}
Law-Smith, J., MacLeod, M., Guillochon, J., Macias, P., \& Ramirez-Ruiz, E.
  2017, The Astrophysical Journal, 841, 132, \dodoi{10.3847/1538-4357/aa6ffb}

\bibitem[{Law-Smith {et~al.}(2020)Law-Smith, Coulter, Guillochon, Mockler, \&
  Ramirez-Ruiz}]{law-smith_stellar_2020}
Law-Smith, J. A.~P., Coulter, D.~A., Guillochon, J., Mockler, B., \&
  Ramirez-Ruiz, E. 2020, The Astrophysical Journal, 905, 141,
  \dodoi{10.3847/1538-4357/abc489}

\bibitem[{Lee {et~al.}(2020)Lee, Offner, Hennebelle, André, Zinnecker,
  Ballesteros-Paredes, Inutsuka, \& Kruijssen}]{lee_origin_2020}
Lee, Y.-N., Offner, S. S.~R., Hennebelle, P., {et~al.} 2020, Space Science
  Reviews, 216, 70, \dodoi{10.1007/s11214-020-00699-2}

\bibitem[{Liu {et~al.}(2023)Liu, Mockler, Ramirez-Ruiz, Yarza, Law-Smith, Naoz,
  Melchor, \& Rose}]{liu_tidal_2023}
Liu, C., Mockler, B., Ramirez-Ruiz, E., {et~al.} 2023, The Astrophysical
  Journal, 944, 184, \dodoi{10.3847/1538-4357/acafe1}

\bibitem[{Lodato {et~al.}(2009{\natexlab{a}})Lodato, King, \&
  Pringle}]{lodato_stellar_2009}
Lodato, G., King, A.~R., \& Pringle, J.~E. 2009{\natexlab{a}}, Monthly Notices
  of the Royal Astronomical Society, 392, 332,
  \dodoi{10.1111/j.1365-2966.2008.14049.x}

\bibitem[{Lodato {et~al.}(2009{\natexlab{b}})Lodato, King, \&
  Pringle}]{Lodato:2009a}
---. 2009{\natexlab{b}}, \mnras, 392, 332.
\newblock
  \url{http://adsabs.harvard.edu/cgi-bin/nph-data_query?bibcode=2009MNRAS.392..332L&link_type=ABSTRACT}

\bibitem[{MacLeod {et~al.}(2012)MacLeod, Guillochon, \&
  Ramirez-Ruiz}]{macleod_tidal_2012}
MacLeod, M., Guillochon, J., \& Ramirez-Ruiz, E. 2012, The Astrophysical
  Journal, 757, 134, \dodoi{10.1088/0004-637X/757/2/134}

\bibitem[{{MacLeod} {et~al.}(2012){MacLeod}, {Guillochon}, \&
  {Ramirez-Ruiz}}]{MacLeod:2012a}
{MacLeod}, M., {Guillochon}, J., \& {Ramirez-Ruiz}, E. 2012, \apj, 757, 134,
  \dodoi{10.1088/0004-637X/757/2/134}

\bibitem[{Mathieu \& Latham(1986)}]{mathieu_spatial_1986}
Mathieu, R.~D., \& Latham, D.~W. 1986, The Astronomical Journal, 92, 1364,
  \dodoi{10.1086/114269}

\bibitem[{Matsuoka {et~al.}(2017)Matsuoka, Nagao, Maiolino, Marconi, Park, \&
  Taniguchi}]{matsuoka_chemical_2017}
Matsuoka, K., Nagao, T., Maiolino, R., {et~al.} 2017, Astronomy and
  Astrophysics, 608, A90, \dodoi{10.1051/0004-6361/201629878}

\bibitem[{Melchor {et~al.}(2023)Melchor, Mockler, Naoz, Rose, \&
  Ramirez-Ruiz}]{melchor_tidal_2023-1}
Melchor, D., Mockler, B., Naoz, S., Rose, S., \& Ramirez-Ruiz, E. 2023, Tidal
  {Disruption} {Events} from the {Combined} {Effects} of {Two}-{Body}
  {Relaxation} and the {Eccentric} {Kozai}-{Lidov} {Mechanism},
  \dodoi{10.48550/arXiv.2306.05472}

\bibitem[{Metzger(2022)}]{metzger_cooling_2022}
Metzger, B.~D. 2022, The Astrophysical Journal, 937, L12,
  \dodoi{10.3847/2041-8213/ac90ba}

\bibitem[{Miller {et~al.}(2023)Miller, Mockler, Ramirez-Ruiz, Draghis, Drake,
  Raymond, Reynolds, Xiang, Yun, \& Zoghbi}]{miller_evidence_2023}
Miller, J.~M., Mockler, B., Ramirez-Ruiz, E., {et~al.} 2023, The Astrophysical
  Journal, 953, L23, \dodoi{10.3847/2041-8213/ace03c}

\bibitem[{{Miller}(2015)}]{Miller:2015a}
{Miller}, M.~C. 2015, \apj, 805, 83, \dodoi{10.1088/0004-637X/805/1/83}

\bibitem[{Milone {et~al.}(2012)Milone, Piotto, Bedin, Aparicio, Anderson,
  Sarajedini, Marino, Moretti, Davies, Chaboyer, Dotter, Hempel, Marín-Franch,
  Majewski, Paust, Reid, Rosenberg, \& Siegel}]{milone_acs_2012}
Milone, A.~P., Piotto, G., Bedin, L.~R., {et~al.} 2012, Astronomy and
  Astrophysics, 540, A16, \dodoi{10.1051/0004-6361/201016384}

\bibitem[{Milosavljević {et~al.}(2006)Milosavljević, Merritt, \&
  Ho}]{milosavljevic_contribution_2006}
Milosavljević, M., Merritt, D., \& Ho, L.~C. 2006, The Astrophysical Journal,
  652, 120, \dodoi{10.1086/508134}

\bibitem[{Mockler {et~al.}(2019)Mockler, Guillochon, \&
  Ramirez-Ruiz}]{mockler_weighing_2019}
Mockler, B., Guillochon, J., \& Ramirez-Ruiz, E. 2019, The Astrophysical
  Journal, 872, 151, \dodoi{10.3847/1538-4357/ab010f}

\bibitem[{Mockler {et~al.}(2023)Mockler, Melchor, Naoz, \&
  Ramirez-Ruiz}]{mockler_uncovering_2023}
Mockler, B., Melchor, D., Naoz, S., \& Ramirez-Ruiz, E. 2023, Uncovering
  {Hidden} {Massive} {Black} {Hole} {Companions} with {Tidal} {Disruption}
  {Events}, \dodoi{10.48550/arXiv.2306.05510}

\bibitem[{Mockler {et~al.}(2022)Mockler, Twum, Auchettl, Dodd, French,
  Law-Smith, \& Ramirez-Ruiz}]{mockler_evidence_2022}
Mockler, B., Twum, A.~A., Auchettl, K., {et~al.} 2022, The Astrophysical
  Journal, 924, 70, \dodoi{10.3847/1538-4357/ac35d5}

\bibitem[{Moe {et~al.}(2019)Moe, Kratter, \& Badenes}]{moe_close_2019}
Moe, M., Kratter, K.~M., \& Badenes, C. 2019, The Astrophysical Journal, 875,
  61, \dodoi{10.3847/1538-4357/ab0d88}

\bibitem[{Nelson {et~al.}(2016)Nelson, van Dokkum, Förster~Schreiber, Franx,
  Brammer, Momcheva, Wuyts, Whitaker, Skelton, Fumagalli, Hayward, Kriek,
  Labbé, Leja, Rix, Tacconi, van~der Wel, van~den Bosch, Oesch, Dickey, \&
  Ulf~Lange}]{nelson_where_2016}
Nelson, E.~J., van Dokkum, P.~G., Förster~Schreiber, N.~M., {et~al.} 2016, The
  Astrophysical Journal, 828, 27, \dodoi{10.3847/0004-637X/828/1/27}

\bibitem[{Nicholl {et~al.}(2020)Nicholl, Wevers, Oates, Alexander, Leloudas,
  Onori, Jerkstrand, Gomez, Campana, Arcavi, Charalampopoulos, Gromadzki,
  Ihanec, Jonker, Lawrence, Mandel, Schulze, Short, Burke, McCully, Hiramatsu,
  Howell, Pellegrino, Abbot, Anderson, Berger, Blanchard, Cannizzaro, Chen,
  Dennefeld, Galbany, González-Gaitán, Hosseinzadeh, Inserra, Irani, Kuin,
  Müller-Bravo, Pineda, Ross, Roy, Smartt, Smith, Tucker, Wyrzykowski, \&
  Young}]{nicholl_outflow_2020}
Nicholl, M., Wevers, T., Oates, S.~R., {et~al.} 2020, Monthly Notices of the
  Royal Astronomical Society, 499, 482, \dodoi{10.1093/mnras/staa2824}

\bibitem[{Parkinson {et~al.}(2022)Parkinson, Knigge, Matthews, Long,
  Higginbottom, Sim, \& Mangham}]{parkinson_optical_2022}
Parkinson, E.~J., Knigge, C., Matthews, J.~H., {et~al.} 2022, Monthly Notices
  of the Royal Astronomical Society, 510, 5426, \dodoi{10.1093/mnras/stac027}

\bibitem[{Paxton {et~al.}(2011)Paxton, Bildsten, Dotter, Herwig, Lesaffre, \&
  Timmes}]{paxton_modules_2011}
Paxton, B., Bildsten, L., Dotter, A., {et~al.} 2011, The Astrophysical Journal
  Supplement Series, 192, 3, \dodoi{10.1088/0067-0049/192/1/3}

\bibitem[{Paxton {et~al.}(2013)Paxton, Cantiello, Arras, Bildsten, Brown,
  Dotter, Mankovich, Montgomery, Stello, Timmes, \&
  Townsend}]{paxton_modules_2013}
Paxton, B., Cantiello, M., Arras, P., {et~al.} 2013, The Astrophysical Journal
  Supplement Series, 208, 4, \dodoi{10.1088/0067-0049/208/1/4}

\bibitem[{Paxton {et~al.}(2015)Paxton, Marchant, Schwab, Bauer, Bildsten,
  Cantiello, Dessart, Farmer, Hu, Langer, Townsend, Townsley, \&
  Timmes}]{paxton_modules_2015}
Paxton, B., Marchant, P., Schwab, J., {et~al.} 2015, The Astrophysical Journal
  Supplement Series, 220, 15, \dodoi{10.1088/0067-0049/220/1/15}

\bibitem[{Paxton {et~al.}(2018)Paxton, Schwab, Bauer, Bildsten, Blinnikov,
  Duffell, Farmer, Goldberg, Marchant, Sorokina, Thoul, Townsend, \&
  Timmes}]{paxton_modules_2018}
Paxton, B., Schwab, J., Bauer, E.~B., {et~al.} 2018, The Astrophysical Journal
  Supplement Series, 234, 34, \dodoi{10.3847/1538-4365/aaa5a8}

\bibitem[{Paxton {et~al.}(2019)Paxton, Smolec, Schwab, Gautschy, Bildsten,
  Cantiello, Dotter, Farmer, Goldberg, Jermyn, Kanbur, Marchant, Thoul,
  Townsend, Wolf, Zhang, \& Timmes}]{paxton_modules_2019}
Paxton, B., Smolec, R., Schwab, J., {et~al.} 2019, The Astrophysical Journal
  Supplement Series, 243, 10, \dodoi{10.3847/1538-4365/ab2241}

\bibitem[{Payne(2021)}]{payne_asassn-14ko_2021}
Payne, A.~V. 2021, The Astrophysical Journal, 20

\bibitem[{Payne {et~al.}(2023)Payne, Auchettl, Shappee, Kochanek, Boyd,
  Holoien, Fausnaugh, Ashall, Hinkle, Vallely, Stanek, \&
  Thompson}]{payne_chandra_2023}
Payne, A.~V., Auchettl, K., Shappee, B.~J., {et~al.} 2023, The Astrophysical
  Journal, 951, 134, \dodoi{10.3847/1538-4357/acd455}

\bibitem[{Price {et~al.}(2024)Price, Liptai, Mandel, Shepherd, Lodato, \&
  Levin}]{price_eddington_2024}
Price, D.~J., Liptai, D., Mandel, I., {et~al.} 2024, Eddington envelopes: {The}
  fate of stars on parabolic orbits tidally disrupted by supermassive black
  holes, \dodoi{10.48550/arXiv.2404.09381}

\bibitem[{Ramirez-Ruiz \& Rosswog(2009{\natexlab{a}})}]{Ramirez-Ruiz:2009a}
Ramirez-Ruiz, E., \& Rosswog, S. 2009{\natexlab{a}}, \apjl, 697, L77.
\newblock
  \url{http://adsabs.harvard.edu/cgi-bin/nph-data_query?bibcode=2009ApJ...697L..77R&link_type=ABSTRACT}

\bibitem[{Ramirez-Ruiz \& Rosswog(2009{\natexlab{b}})}]{ramirez-ruiz_star_2009}
---. 2009{\natexlab{b}}, The Astrophysical Journal, 697, L77,
  \dodoi{10.1088/0004-637X/697/2/L77}

\bibitem[{Rees(1988)}]{Rees:1988a}
Rees, M.~J. 1988, \nat, 333, 523.
\newblock \url{http://adsabs.harvard.edu/abs/1988Natur.333..523R}

\bibitem[{Rosswog {et~al.}(2009)Rosswog, Ramirez-Ruiz, \&
  Hix}]{rosswog_tidal_2009}
Rosswog, S., Ramirez-Ruiz, E., \& Hix, W.~R. 2009, The Astrophysical Journal,
  695, 404, \dodoi{10.1088/0004-637X/695/1/404}

\bibitem[{{Roth} \& {Kasen}(2018)}]{Roth:2018}
{Roth}, N., \& {Kasen}, D. 2018, \apj, 855, 54,
  \dodoi{10.3847/1538-4357/aaaec6}

\bibitem[{Roth {et~al.}(2016)Roth, Kasen, Guillochon, \&
  Ramirez-Ruiz}]{roth_x-ray_2016}
Roth, N., Kasen, D., Guillochon, J., \& Ramirez-Ruiz, E. 2016, The
  Astrophysical Journal, 827, 3, \dodoi{10.3847/0004-637X/827/1/3}

\bibitem[{Ryu {et~al.}(2020)Ryu, Krolik, \& Piran}]{ryu_measuring_2020}
Ryu, T., Krolik, J., \& Piran, T. 2020, The Astrophysical Journal, 904, 73,
  \dodoi{10.3847/1538-4357/abbf4d}

\bibitem[{{Ryu} {et~al.}(2020){Ryu}, {Krolik}, \& {Piran}}]{Ryu:2020a}
{Ryu}, T., {Krolik}, J., \& {Piran}, T. 2020, \apj, 904, 73,
  \dodoi{10.3847/1538-4357/abbf4d}

\bibitem[{Ryu {et~al.}(2020)Ryu, Krolik, Piran, \& Noble}]{ryu_tidal_2020}
Ryu, T., Krolik, J., Piran, T., \& Noble, S.~C. 2020, The Astrophysical
  Journal, 904, 98, \dodoi{10.3847/1538-4357/abb3cf}

\bibitem[{Somalwar {et~al.}(2023)Somalwar, Ravi, Yao, Guolo, Graham,
  Hammerstein, Lu, Nicholl, Sharma, Stein, van Velzen, Bellm, Coughlin, Groom,
  Masci, \& Riddle}]{somalwar_first_2023}
Somalwar, J.~J., Ravi, V., Yao, Y., {et~al.} 2023, The first systematically
  identified repeating partial tidal disruption event,
  \dodoi{10.48550/arXiv.2310.03782}

\bibitem[{Steinberg \& Stone(2022)}]{steinberg_origins_2022}
Steinberg, E., \& Stone, N.~C. 2022, The {Origins} of {Peak} {Light} in {Tidal}
  {Disruption} {Events}, Tech. rep.
\newblock \url{https://ui.adsabs.harvard.edu/abs/2022arXiv220610641S}

\bibitem[{Stone \& Metzger(2016)}]{stone_rates_2016}
Stone, N.~C., \& Metzger, B.~D. 2016, 25

\bibitem[{Tejeda {et~al.}(2017)Tejeda, Gafton, Rosswog, \&
  Miller}]{tejeda_tidal_2017}
Tejeda, E., Gafton, E., Rosswog, S., \& Miller, J.~C. 2017, Monthly Notices of
  the Royal Astronomical Society, 469, 4483, \dodoi{10.1093/mnras/stx1089}

\bibitem[{Wang \& Merritt(2004)}]{wang_revised_2004}
Wang, J., \& Merritt, D. 2004, The Astrophysical Journal, 600, 149,
  \dodoi{10.1086/379767}

\bibitem[{Wen {et~al.}(2020)Wen, Jonker, Stone, Zabludoff, \&
  Psaltis}]{wen_continuum-fitting_2020}
Wen, S., Jonker, P.~G., Stone, N.~C., Zabludoff, A.~I., \& Psaltis, D. 2020,
  The Astrophysical Journal, 897, 80, \dodoi{10.3847/1538-4357/ab9817}

\bibitem[{Wevers {et~al.}(2023)Wevers, Guolo, Pasham, Coughlin, Tombesi, Yao,
  \& Gezari}]{wevers_delayed_2023}
Wevers, T., Guolo, M., Pasham, D.~R., {et~al.} 2023, Delayed {X}-ray
  brightening accompanied by variable ionized absorption following a tidal
  disruption event, \dodoi{10.48550/arXiv.2311.09371}

\bibitem[{Yang {et~al.}(2017{\natexlab{a}})Yang, Wang, Ferland, Dou, Zhou,
  Jiang, \& Sheng}]{Yang:2017}
Yang, C., Wang, T., Ferland, G.~J., {et~al.} 2017{\natexlab{a}}, The
  Astrophysical Journal, 846, 150, \dodoi{10.3847/1538-4357/aa8598}

\bibitem[{Yang {et~al.}(2017{\natexlab{b}})Yang, Wang, Ferland, Dou, Zhou,
  Jiang, \& Sheng}]{yang_carbon_2017}
---. 2017{\natexlab{b}}, The Astrophysical Journal, 846, 150,
  \dodoi{10.3847/1538-4357/aa8598}

\end{thebibliography}
\bibliographystyle{aasjournal}

\end{document}